%% file: main.tex
\documentclass[10pt,conference]{IEEEtran}

\usepackage{cite}
\usepackage{amsmath,amssymb,amsfonts}
\usepackage{algorithmic}
\usepackage{graphicx}
\usepackage{textcomp}
\usepackage{xcolor}
\usepackage[hyphens]{url}

\def\BibTeX{{\rm B\kern-.05em{\sc i\kern-.025em b}\kern-.08em
    T\kern-.1667em\lower.7ex\hbox{E}\kern-.125emX}}

\usepackage{multirow}
\usepackage{listings}
\usepackage{hyperref}

\usepackage{tikz}
\usetikzlibrary{positioning, decorations.pathreplacing}
\usepackage{subcaption}

\usepackage[normalem]{ulem}
\usepackage[draft]{fixme}

\usepackage{cleveref}

\newcommand{\runinsec}[1]{\noindent\textbf{#1:} }

\title{A Storage-Effective BTB Organization for Servers} 

\author{\IEEEauthorblockN{Truls Asheim}
\IEEEauthorblockA{NTNU, Norway}
\and
\IEEEauthorblockN{Boris Grot}
\IEEEauthorblockA{University of Edinburgh, UK}
\and
\IEEEauthorblockN{Rakesh Kumar}
\IEEEauthorblockA{NTNU, Norway}
}

\begin{document}

\maketitle
\thispagestyle{plain}
\pagestyle{plain}


\input{0_abstract.tex}

\input{1_intro}
\input{2_background}

\input{3_targetDistanceAnalysis}
\input{4_btbOrganizations}
\input{5_design}
\input{6_eval}
\input{7_related}
\input{8_conclusion}

\section*{Acknowledgements}
This work is partially supported through the Research Council of Norway (NFR) grant 302279 to NTNU.



\bibliographystyle{IEEEtranS}
\bibliography{refs}

\appendices

\input{9_appendix}

\end{document}

%% file: 0_abstract.tex
\begin{abstract}

Many contemporary applications feature multi-megabyte instruction footprints that overwhelm the capacity of branch target buffers (BTB) and instruction caches (L1-I), causing frequent front-end stalls that inevitably hurt performance. BTB capacity is crucial for performance as a sufficiently large BTB enables the front-end to accurately resolve the upcoming execution path and steer instruction fetch appropriately. Moreover, it also enables highly effective fetch-directed instruction prefetching that can eliminate a large portion L1-I misses. For these reasons, commercial processors allocate vast amounts of storage capacity to BTBs. 

This work aims to reduce BTB storage requirements by optimizing the organization of BTB entries. Our key insight is that storing branch target offsets, instead of full or compressed targets, can drastically reduce BTB storage cost as the vast majority of dynamic branches have short offsets requiring just a handful of bits to encode. Based on this insight, we size the ways of a set associative BTB to hold different number of target offset bits such that each way stores offsets within a particular range. Doing so enables a dramatic reduction in storage for target addresses. Our final design, called BTB-X, uses an 8-way set associative BTB with differently sized ways that enables it to track about 2.24x more branches than a conventional BTB and 1.3x more branches than a storage-optimized state-of-the-art BTB organization, called PDede, with the same storage budget.
\end{abstract} 

%% file: 1_intro.tex
\section{Introduction}
\label{sec:intro}

Contemporary server applications feature massive instruction footprints stemming from deeply layered software stacks. These footprints may far exceed the capacity of the branch target buffer (BTB) and instruction cache (L1-I), resulting in the so-called front-end bottleneck. BTB misses may lead to wrong-path execution, triggering a pipeline flush when misspeculation is detected. Such pipeline flushes not only throw away tens of cycles of work but also expose the fill latency of the pipeline. Similarly, L1-I misses cause the core front-end to stall for tens of cycles while the miss is being served from lower-level caches. 

BTB stands at the center of a high-performance core front end for three key reasons: it determines the instruction stream to be fetched, it identifies branches for the branch direction predictor, and it affects the L1-I hit rate. Specifically, by identifying control flow divergences, the BTB ensures that the branch direction predictor can make predictions for upcoming conditional branches. For predicted-taken and unconditional branches, the BTB supplies targets to which instruction fetch should be redirected. Finally, the BTB together with the direction predictor enables an important class of instruction prefetchers called fetch-directed instruction prefetchers (FDIP)~\cite{fdip, boomerang, shotgun, shotgunTOCS}, which rely on the BTB to discover L1-I prefetch candidates. 

Considering the criticality of capturing the large branch working sets of modern workloads, commercial CPUs feature BTBs with colossal capacities, a trend also observed by ~\cite{rebase}. With each BTB entry potentially requiring 8 bytes or more (Section~\ref{sec:background}), BTB storage costs can easily reach into tens and even hundreds of KBs. Indeed, the Samsung Exynos M6 mobile processor allocates a staggering 529KB of on-chip storage to BTBs~\cite{exynos}. Not only the BTB storage cost is high, it is increasing at a rapid pace. For example, the Samsung Exynos BTB storage budget increased nearly six fold (98.9KB to 561.5KB) from M2 to M6, over a period of about eight years~\cite{exynos}. While such massive BTBs are effective at capturing branch working sets, they do so at staggering area costs.

\begin{figure}
\centering
\includegraphics[width=\columnwidth]{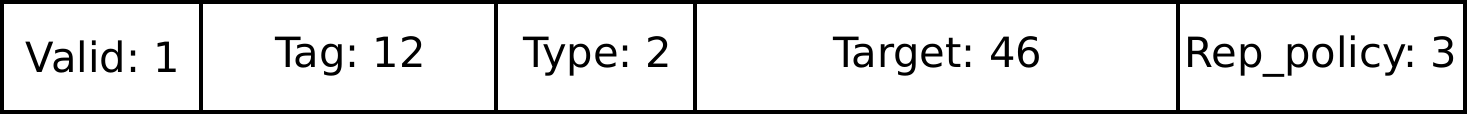}
\caption {Composition of an entry in conventional BTB. The numbers are the number of bits required to encode each field.}
\label{fig:conv-btb}
\end{figure}

To reduce the BTB storage cost, prior work~\cite{DUPN, ittage, pdede} has focused on compressing the branch targets as they account for the majority of BTB storage budget as shown in Figure~\ref{fig:conv-btb}. Concretely, Seznec ~\cite{DUPN, ittage} observes that all branch targets within a page share the same page number and BTB storage requirements can be significantly reduced by storing the page number only once per page instead of once per target. To exploit this observation, he partitions the BTB in two structures, Main-BTB and Page-BTB, each storing different portions of branch targets. The Main-BTB stores the page offset and a pointer to the Page-BTB entry that stores the page number. The state-of-the-art BTB organization, PDede~\cite{pdede}, further observes that target addresses span significantly fewer \emph{regions} than pages, where a region is a group of contiguous pages. Therefore, it partitions the BTB even further and introduces a Region-BTB that lowers page number storage cost as the region number for all pages inside a region is stored only once. By storing page/region numbers only once for all branches in a page/region, these BTBs avoid information duplication, thus reducing storage requirements.

Though, these designs significantly reduce BTB storage requirements, they introduce several complexities that increase their access latency and power requirements. First, these designs introduce a level of indirection, i.e., on a BTB access, Main-BTB is accessed first to get the pointers to the Page-BTB and Region-BTB and only then these BTBs can be accessed. This sequential access, Main-BTB followed by Page/Region-BTBs, increases the overall BTB access latency which either requires a two-cycle BTB lookup or a longer clock period. Both of these alternatives incur a performance penalty. Second, on allocating a new BTB entry, Page-BTB and Region-BTB need to be searched to check if the page/region number for the target address is already present or not. As the page/region number can be anywhere in Page/Region-BTB, a fully-associative associative search is required~\cite{ittage} which increases BTB power requirements. An alternative is to restrict the locations where a page/region number can be stored in Page/Region-BTB~\cite{pdede}; however, it increase the likelihood of conflict misses.

This work seeks to reduce BTB storage requirements without increasing BTB complexity. To that end, we propose to store target offsets, defined as delta between the address of the branch instruction and that of its target, instead of full or compressed (i.e., page offset, page number, and region number) targets. Our key insight is that target offsets are unevenly distributed but tend to require significantly fewer bits to represent than full and even compressed target addresses. Our analysis reveals that 54\% of dynamic branches require only 6 bits or fewer for offset encoding, while a meager 1\% of branches need 25 bits or more to store their offsets.

Based on this insight, we propose to store target offsets in the BTB rather than compressed or full target addresses, which can be up to 64 bits long depending on the size of virtual address space. To accommodate the varied distribution of target offsets, we size different ways of a set associative BTB to hold different number of offset bits such that each way stores only those branches whose target offsets can be encoded with a certain number of bits. In doing so, we not only significantly reduce BTB storage requirements but also avoid the complexities, indirection and fully-associative searches of the state-of-the-art BTB designs.

This paper introduces BTB-X, a simple yet highly storage-effective BTB organization, that incarnates our idea of storing target offsets. BTB-X is a set associative BTB with its ways sized to store different sized target offsets. Our evaluation shows that BTB-X can track about 2.24x more branches than a conventional BTB storing full targets and about 1.3x more branches than PDede, a state-of-the-art BTB organization, with the same storage budget. Conversely, BTB-X can accommodate the same number of branches as conventional BTB and PDede while requiring 2.24x and 1.3x less storage. This work makes the following key contributions:

\begin{itemize}
    \item We show that storing branch target offsets, instead of full or compressed target addresses, can provide drastic BTB storage savings because about 54\% of branches require only 6 bits or fewer to encode their offsets. A further 22\% of branches require between 7 and 10 bits. 
    \item We show that the target offset sizes are unevenly distributed with 0-6 bits, 7-10 bits, and 11-25 bits required to encode the offsets of 54\%, 22\% and 23\% of branches respectively. Therefore, a single size offset field cannot provide storage optimal solution.
    \item We introduce BTB-X, a simple and highly storage-effective BTB organization, that stores target offsets instead of targets themselves. Furthermore, BTB-X ways are sized to hold different sized target offsets.  
    \item We demonstrate that, with the same storage budget, BTB-X can accommodate about 2.24x and 1.3x more branches than a conventional BTB and PDede, a state-of-the-art BTB. Our evaluation further shows that BTB-X outperforms the conventional BTB even when provisioned with just half the storage budget.
\end{itemize}

%% file: 2_background.tex
\section{Background and Motivation}
\label{sec:background}

Branch prediction unit predicts the program control flow and supplies a stream of instruction addresses/program counters (PCs) on the predicted path to the fetch unit which fetches the corresponding instructions to feed the rest of the core. As branch instructions disturb the otherwise sequential control flow, the branch prediction unit needs to identify them to predict the upcoming control flow. However, whether an instruction is a branch or not can only be determined after it has been fetched and decoded. To avoid the latency of fetching and decoding instructions before generating next PCs, the branch prediction unit employs a special hardware structure, called branch target buffer (BTB), to identify branch instructions solely from their PCs before the instructions themselves are even fetched.

\subsection{Branch Target Buffer (BTB)}
Figure~\ref{fig:conv-btb} presents the conventional BTB organization. Each BTB entry consists of \textit{valid}, \textit{tag}, \textit{type}, \textit{target}, and \textit{rep\_policy} fields. Figure~\ref{fig:conv-btb} also shows the typical number of bits needed for these fields. The \textit{tag} field usually stores only a partial tag, which is generated by hashing the full tag, to reduce storage cost while introducing minimal aliasing. The number of bits for \textit{target} field depends on the size of virtual address space and instruction set architecture (ISA). We assume a 48-bit virtual address space and ARMv8 ISA to calculate target field size in Figure~\ref{fig:conv-btb}. As ARMv8 instructions are always 32-bits and 4-byte aligned, the least significant two bits of a PC are always zeros. Therefore, we only need 46-bits for the \textit{target} field. The \textit{valid} bit indicates whether the entry contains valid information or not, while \textit{rep\_policy} bits choose one of the existing branches for eviction when a new branch is inserted in the BTB.

To check whether a PC corresponds to a branch instruction, the BTB is indexed, i.e. accessed, with the low order PC bits. The high order PC bits are hashed, using the same function that is used to generate partial tags, and compared to the \textit{tag} field of the indexed BTB entry. A match indicates that the PC belongs to a branch. 

A branch instruction simply implies the presence of a potential control flow divergence point in program execution. However, whether or not the divergence actually happens depends on the type of branch, i.e. call, return, conditional, or unconditional branch, which is stored in the \textit{type} field of a BTB entry. Call, return, and unconditional branches always cause control flow divergence as they are always \textit{taken}. Conditional branches, in contrast, are not always taken and a direction predictor is used to predict their direction. 

If a branch is predicted to be taken, the \textit{target} field in the BTB entry provides the address for the next instruction, except for returns. This is because the return address is call-site dependent and a given function can be called from different call sites. Therefore, a return address stack (RAS) is typically employed to record return addresses at call-sites. On a function call, the call instruction pushes the return address to RAS, which is later popped by the corresponding return instruction.

\subsection{The cost of a BTB miss}
A BTB miss for a branch instruction means that the branch is undetected and the front-end continues to fetch instructions sequentially. Whether or not the sequential path is the correct one depends on the actual direction of the missed branch. Unless the missed branch is a conditional branch that is not taken, the sequential path is incorrect. When the wrong path is eventually detected by the core, all the instructions after the branch that missed in the BTB are flushed, fetch is redirected to the branch target and pipeline is filled with correct-path instructions. BTB misses are thus highly deleterious to performance as they result in a loss of tens of cycles of work and expose the pipeline fill latency. 

\begin{figure}
\centering
\includegraphics[width=.9\columnwidth, trim=0 0 0 0, clip]{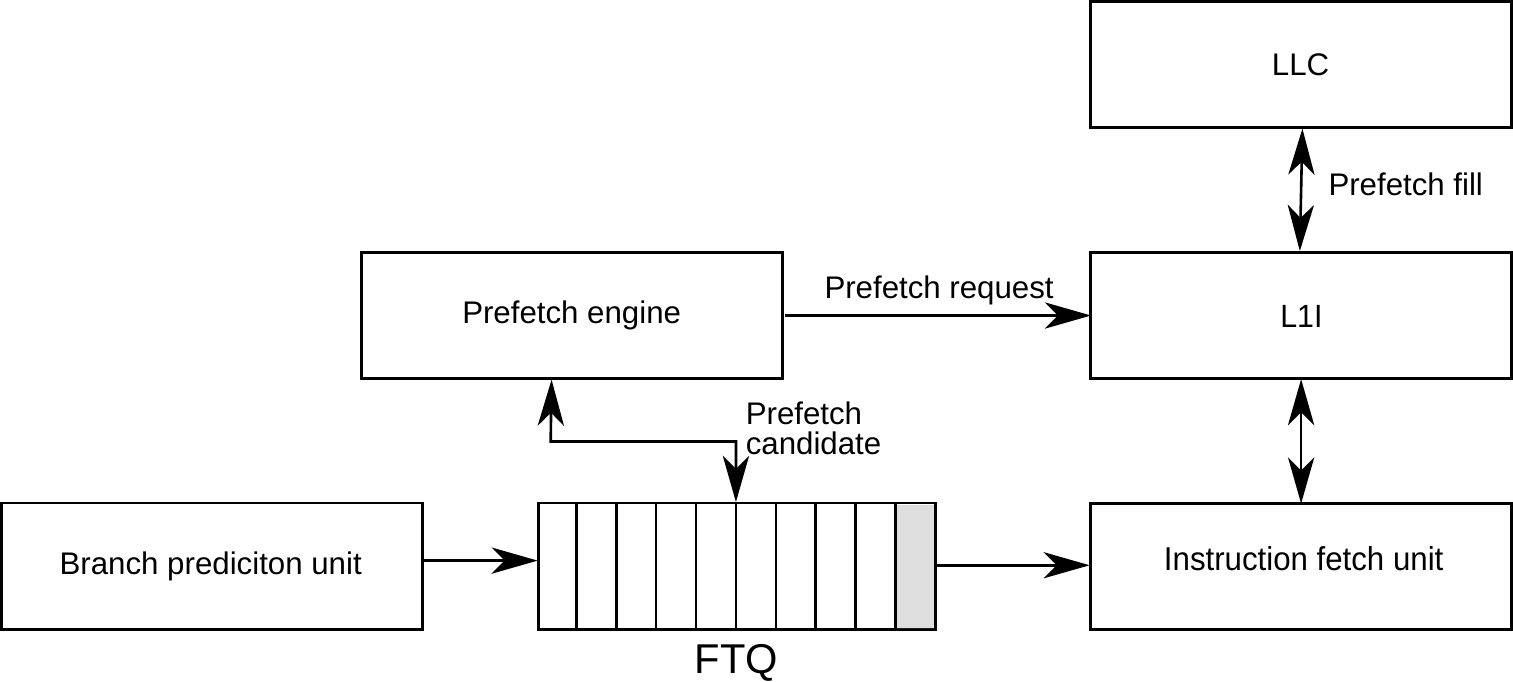}
\caption{FDIP microarchitecture}
\label{fig:fdip}
\end{figure}

\subsection{BTB's role in instruction prefetching}
Fetch-directed instruction prefetchers are a class of powerful L1-I prefetchers that intrinsically rely on BTB to identify prefetch candidates. 
These prefetchers are highly effective and, when coupled with a sufficiently large BTB, outperform the winner of the recently-concluded Instruction Prefetching Championship~\cite{ipc1} and approach the performance of an ideal L1-I, as reported by Ishii et al.~\cite{rebase}. Variants of these prefetchers have been adopted in commercial products, for example in IBM z15~\cite{IBMz15HotChips}, ARM Neoverse N1~\cite{neoverse} etc.

\Cref{fig:fdip} shows a canonical organization of a fetch-directed instruction prefetcher (FDIP)~\cite{fdip}. 
As originally proposed, FDIP decouples the branch-prediction unit and the fetch engine via the {\em fetch target queue (FTQ)}. This decoupling allows the branch prediction unit to run ahead of the fetch engine and discover prefetch candidates by predicting the control flow far into the future. With FDIP, each cycle, the branch prediction unit identifies and predicts branches to anticipate upcoming execution path and inserts corresponding instruction addresses into the FTQ. Consequently, the FTQ contains a stream of anticipated instruction addresses to be fetched by the core. The prefetch engine scans the FTQ to identify prefetch candidates and issue prefetch requests. 

For FDIP to be effective, the BTB needs to accommodate the branch working set, otherwise frequent BTB misses will cause FDIP to prefetch the wrong path as FTQ will be filled with wrong path instruction addresses. This is one of the key reasons why commercial processors deploy massive BTBs, as also observed by~\cite{rebase}. These massive BTBs incur astronomical storage overheads. Also, not only the BTB storage overhead is high, it is increasing at a rapid pace. For example, Table~\ref{tab:btbcap}, presents the BTB storage cost in several generations of Samsung Exynos processors. As the table shows, the BTB storage cost nearly doubled in each generation, except between M4 and M5. Overall, the storage cost increased nearly six fold (98.9KB to 561.5KB) from M1 to M6, over a period of about eight years~\cite{exynos}.

As the instruction footprints of server applications continue to expand, a trend also reflected in Google Web Search workload whose instruction footprint is growing at annualized rate of 27\%~\cite{profileWarehouse}, the BTB sizes and their storage overheads are destined to increase in future. Therefore, there is an urgent need to investigate storage-effective BTB organizations to combat the front-end bottleneck without necessitating prohibitive area budgets.

\begin{table}[t]
  \centering
  \caption{\label{tab:btbcap} BTB storage cost in Samsung Exynos processors}
  \begin{tabular}{ll}\hline
    \textbf{CPU} & \textbf{BTB Storage} \\\hline
    M1/M2 & 98.9KB \\
    M3 & 175.8KB \\
    M4 & 288.0KB \\
    M5 & 310.8KB \\
    M6 & 561.5KB \\\hline
  \end{tabular}
\end{table}

%% file: 3_targetDistanceAnalysis.tex
\section{Branch Target Distance Analysis}
\label{sec:analysis}

\begin{figure}[b]
  \sffamily
  \centering
  \resizebox{\columnwidth}{!}{
  \begin{tabular}{p{2.2cm}|rlllllllllll}
    Bit position      & 48 & ... & 9 & 8 & 7 & 6 & 5 & 4 & 3 & 2 & 1\\\hline
    Branch PC         &  0 & ... & 1 & 0 & 1 & 1 & 0 & 1 & 0 & 0 & 0\\
    Branch Target     &  0 & ... & 1 & 0 & 1 & 1 & 1 & 1 & 0 & 0 & 0\\
    Target Offset     &    & ... &   &   &   &   & 1 & 1 & 0 & 0 & 0\\
  \end{tabular}}
  \caption{\label{fig:concat} Branch target offset example}
\end{figure}

\begin{figure*}[t!]
    \centering
    \includegraphics[width=0.7\textwidth, trim=60 280 50 320, clip]{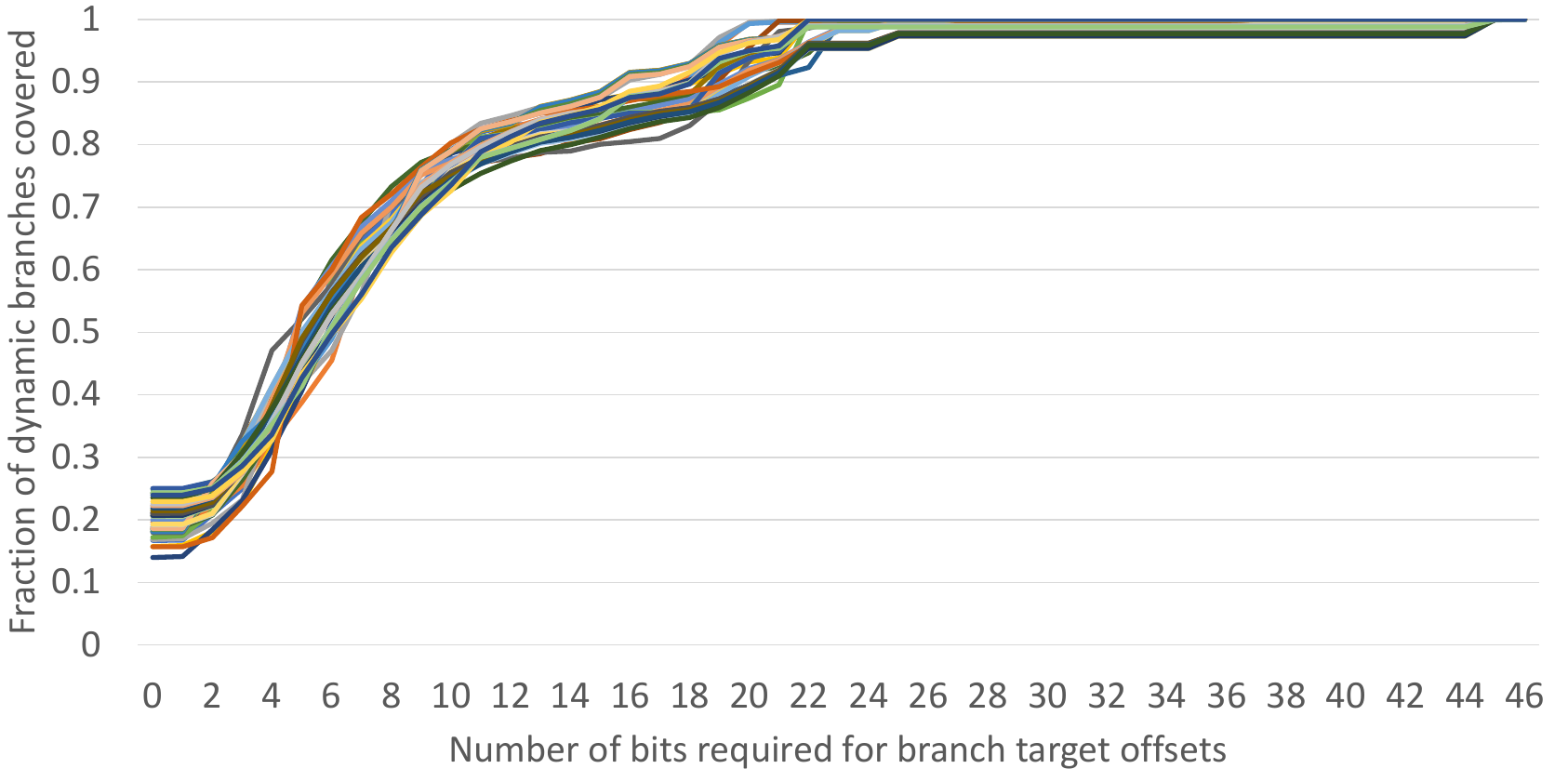}
    \caption{Distribution of branch target offsets in different workloads.}
    \label{fig:offsets}
\end{figure*}

The storage cost of branch targets accounts for a major fraction of BTB storage requirements. For example, in a conventional BTB, as depicted in Figure~\ref{fig:conv-btb}, the target field accounts for about 72\% (46 of 64 bits) of the total BTB storage requirements. We analyze the number of bits required for branch target \emph{offsets} to assess if storing the offsets, instead of the full or compressed targets, can reduce BTB storage requirements. We define the target offset as the \textit{n} least significant bits of target address, with \textit{n} being the position of most significant bit that differs among branch PC and target. As an example, for the branch PC and target shown in Figure~\ref{fig:concat}, the most significant bit that differs among them is at position five, whereas all bits at positions higher than five are same. Therefore, the target offset for this branch PC and target pair is `11000', i.e., the target bits from position 5 to 1. Also, as our modelled ARM v8 ISA aligns instructions at 4-byte boundaries, the two least significant bits of a target are always zeros. Therefore, we only need to store `110' as offset in the BTB.

An advantage of defining an offset as \textit{n} lower order target bits, instead of the numerical distance between branch PC and target (i.e., target - PC), is that the full target can be recovered by simply concatenating the shifted branch PC with the offset retrieved from BTB. In contrast, using numerical distance as offset would require a 48-bit adder to recover the full target from offset. 

Figure~\ref{fig:offsets} plots the distribution of branch target offsets in the branch working sets of our workloads. The data includes both conditional and unconditional branches; hence, it comprehensively covers the full branch working set. The X-axis shows the number of bits required to store offsets, while the Y-axis plots the fraction of dynamic branches covered.

As the figure shows, short offsets dominate the distribution with 54\% of branches requiring only six bits or fewer for their offsets. A further 22\% of branches only require between 7 and 10-bits to represent their offsets. The reason why such a high fraction of offsets is short is that conditional branches dominate the dynamic branch working set, and they tend to have short offsets~\cite{boomerang}. This is because conditional branches generally guide the control flow only inside a function; meanwhile, software engineering principles favor small functions, thus restricting conditional branch target offsets to short distances. Furthermore, as discussed in Section~\ref{sec:background}, return instructions get their target from RAS, thus they do not need to store any target bits in BTBs. Therefore, for the purpose of this analysis, we assume 0-bit offsets for return instructions.

Perhaps surprisingly, Figure~\ref{fig:offsets} also shows that very few branches require a large number of bits for their offset. Indeed, a meagre 1\% of branches requires more than 25 bits for their offsets. The sum of these results indicates that reserving space for the full 46-bit target address results in an appalling under-utilization of BTB storage, since 99\% of branches need at most half the number of bits needed to represent the full target address if offsets are used instead.

We gain two key insights from this analysis:

\runinsec{Key Insight 1} The targets of most branches lie relatively close in the virtual address space to the branch itself. As a result, storing the {\em distance} to the target, in the form of an offset from the branch instruction can provide drastic storage savings. 

\runinsec{Key Insight 2} The target offset sizes are unevenly distributed with 0-6 bits, 7-10 bits, and 11-25 bits required to encode the offsets of 54\%, 22\% and 23\% of branches respectively. Therefore, a single size offset field cannot provide storage optimal solution.

%% file: 4_btbOrganizations.tex
\section{State-of-the-art BTBs and Their Limitations}
\label{sec:btbOrgs}

Prior work has proposed several BTB organizations that aim to reduce the storage cost by compressing branch targets. This section presents the most representative BTB organizations and analyzes their limitations.

\subsection{Reduced BTB:} Seznec~\cite{DUPN} made a critical observation that all branch targets within a page share the same page number and only differ in page offsets. Thus, storing full target addresses in a BTB results in massive duplication of page numbers and wastage of storage capacity. To eliminate this duplication, Seznec proposed Reduced BTB (R-BTB), a variant of which was also used in ITTAGE~\cite{ittage}. The key innovation of R-BTB is to store a pointer to the page number rather than storing the page number itself in BTB. 

\begin{figure}
\centering
\includegraphics[width=.9\columnwidth, trim=0 0 0 0, clip]{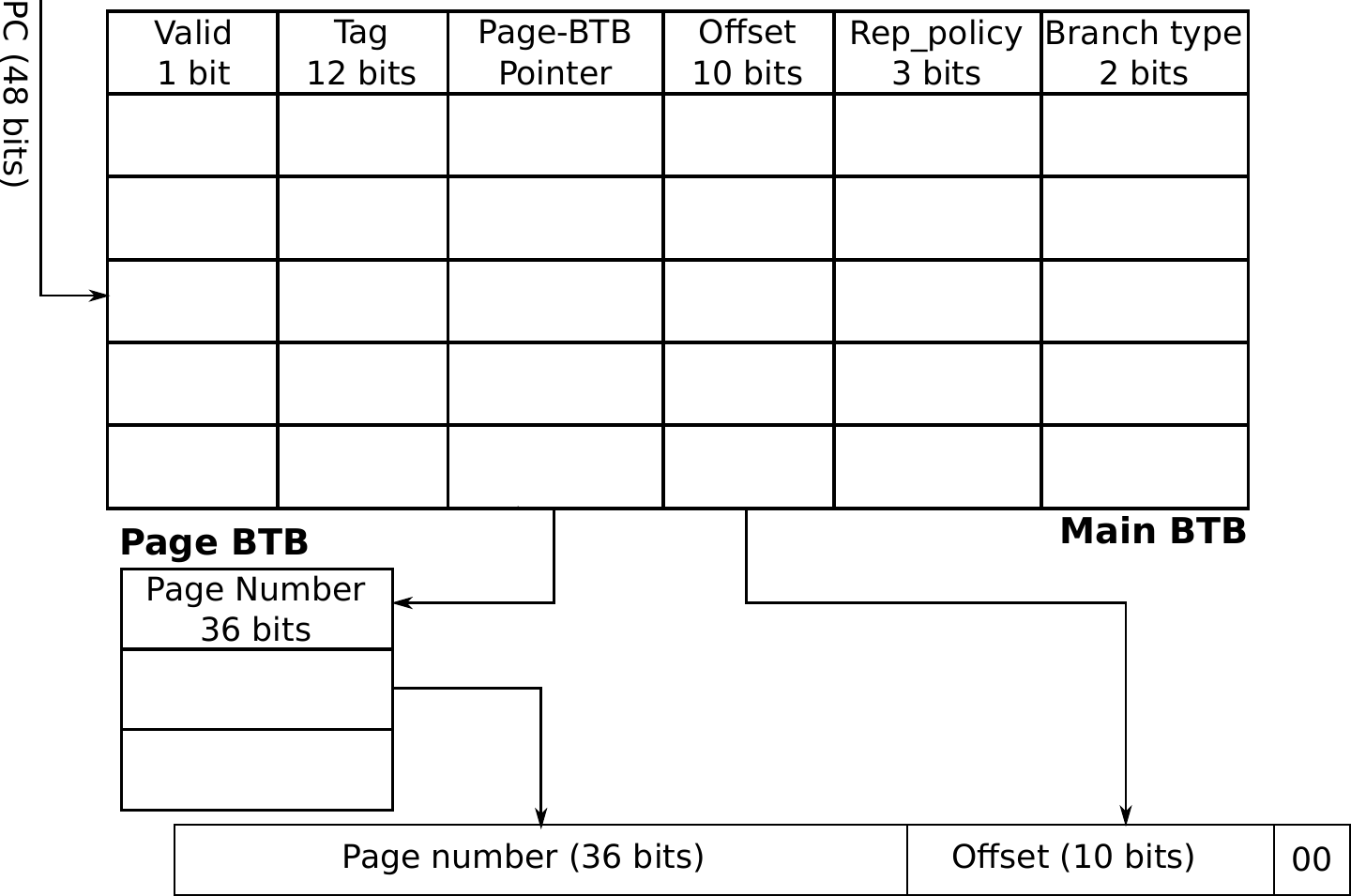}
\caption{Reduced BTB Organization.}
\label{fig:rbtb}
\end{figure}

Figure~\ref{fig:rbtb} presents the logical organization of R-BTB and the composition of its entries. R-BTB is composed of two partitions: Main-BTB and Page-BTB. For each branch target, apart from page offset, Main-BTB stores a pointer to the page number. The page number itself is stored in Page-BTB. If two or more branches have their targets in the same page, their Main-BTB entries will hold pointer to the same Page-BTB entry. As the number of pages is significantly smaller than number of branch targets, fewer bits are needed to hold Page-BTB pointers than page numbers themselves. Consequently, by storing a page number only once in Page-BTB, R-BTB avoids duplication and reduces storage requirements. 

\subsection{PDede:} 
\label{sec:pdedearch}
PDede~\cite{pdede} is the state-of-the-art BTB organization that comes with three different variants. Figure~\ref{fig:pdede} depicts the most storage effective and best performing PDede variant, called PDede-Multi Entry Size. It improves over R-BTB in two aspects. First, it reduces the cost of storing page numbers in the Page-BTB. PDede observes that server applications, due to their large instruction footprints, touch a large number of pages thus increasing Page-BTB storage requirements. They further observe that, as different libraries get dynamically mapped to different locations in address space,  the pages tend to form spatial \emph{regions}, where a region consists of multiple contiguous pages. Just like branch targets inside a page share the same page number, the page numbers inside a region share the same region number. To eliminate the duplication of region numbers, as shown in Figure~\ref{fig:pdede} PDede introduces a Region-BTB which stores the region number while Main-BTB stores a pointer to it just like it stores a pointer to page BTB. 

\begin{figure}
\centering
\includegraphics[width=.9\columnwidth, trim=0 0 0 0, clip]{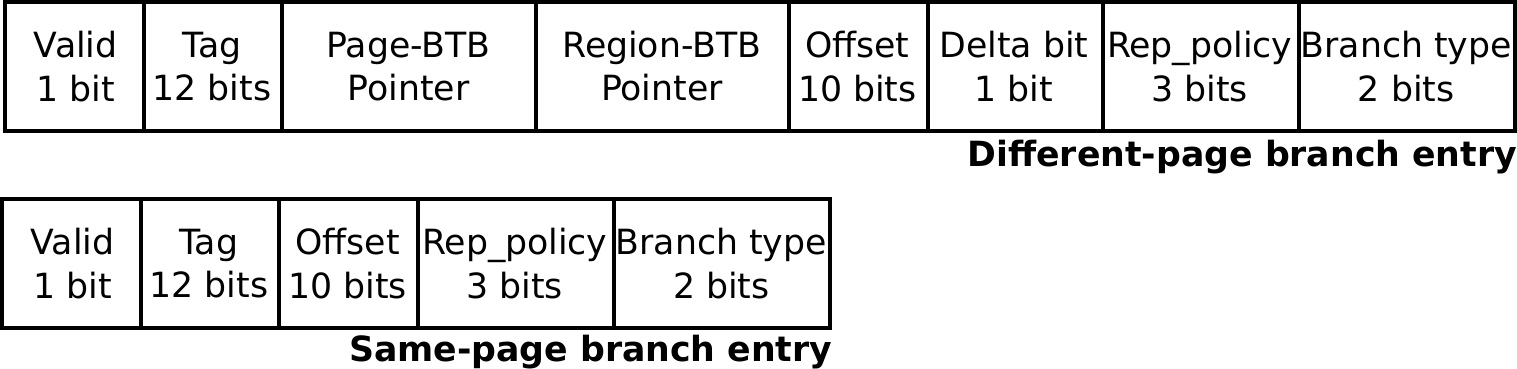}
\caption{Different- and same-page PDede entry composition.}
\label{fig:pdedeEntry}
\end{figure}

Second, for the same-page branches, i.e., when the branch and its target are in the same page, PDede does not store page/region numbers as they can be recovered from branch PC. PDede reserves half of the ways in a set associate BTB for same-page branches. As the ways reserved for same-page branches do not need to store Page-BTB and Region-BTB pointers, as shown in Figure~\ref{fig:pdedeEntry}, PDede achieves additional storage savings.

\subsection{Limitations of the state-of-the-art:} Though R-BTB and PDede achieve significant storage saving by avoiding page and region number duplication, they increase BTB complexity by introducing a level of indirection and associative searches in Page- and Region-BTB. These complexities lead to increased access latency and power requirements. In addition, the state-of-the-art BTB designs are suboptimal in utilizing the available storage budget.

\begin{figure}
\centering
\includegraphics[width=.9\columnwidth, trim=0 0 0 10, clip]{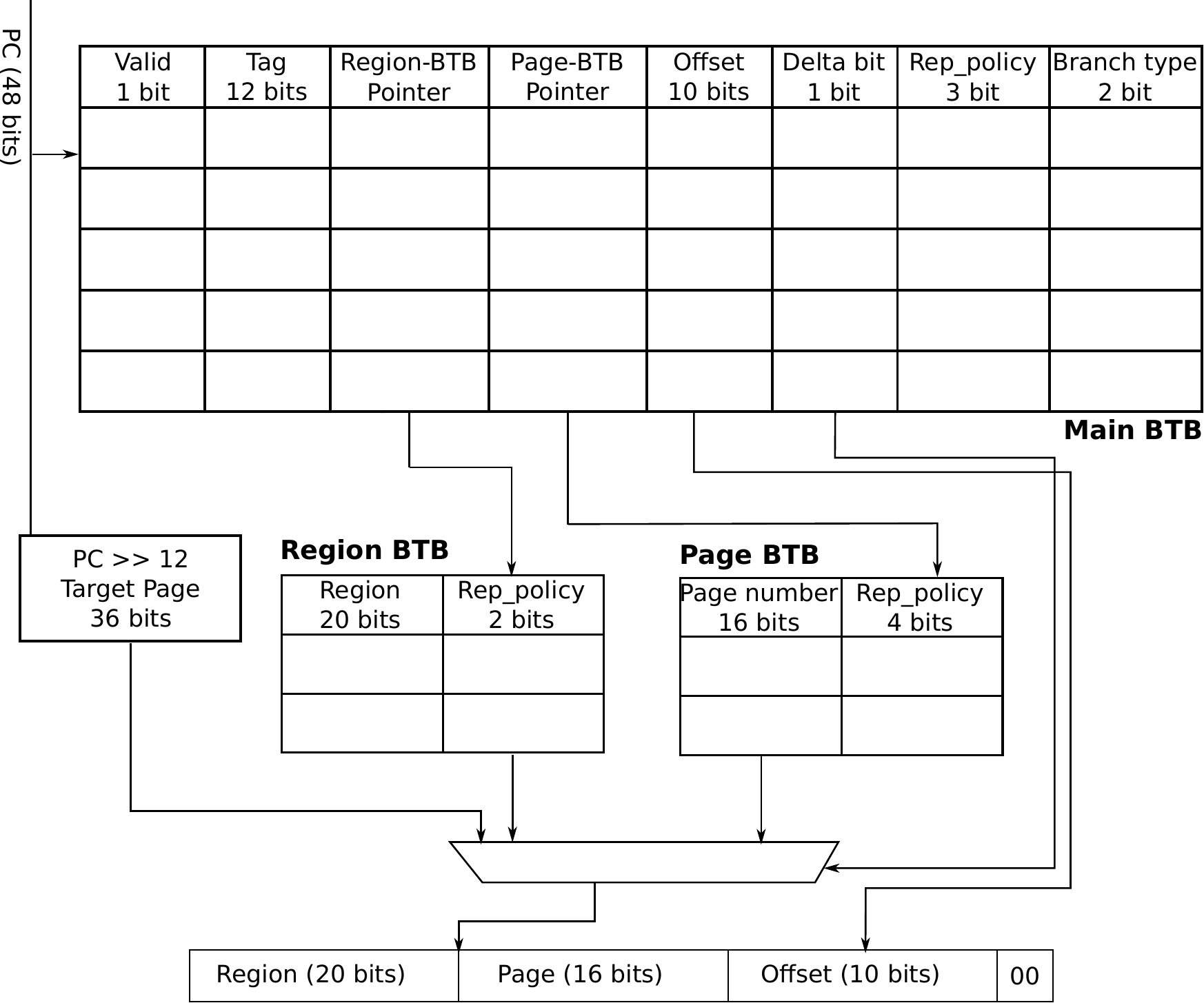}
\caption{PDede BTB Organization.}
\label{fig:pdede}
\end{figure}

\runinsec{Indirection} As Figures~\ref{fig:rbtb} and~\ref{fig:pdede} show, the access to Main-BTB only provides a part of the target address, i.e., page offset. The other parts have to be retrieved from Page-BTB and Region-BTB. Also, Page- and Region-BTB cannot be accessed in parallel with Main-BTB because the Main-BTB access provide the pointers to them. As a result, the sequential Main-BTB and Page-/Region-BTB accesses increase the overall BTB access latency. This additional latency either enforces a two-cycle BTB lookup or necessitates a longer clock period. Both of these alternatives are detrimental to performance.  PDede does avoid this indirection penalty to some extent because the same-page branches do not need to access Page-/Region-BTB rather they get their page and region number from the branch PC itself. However, the different-page branches, i.e., where branch PC and target address lie in different pages, do need to pay the indirection penalty.

\begin{figure*}
    \centering
    \includegraphics[width=0.8\textwidth]{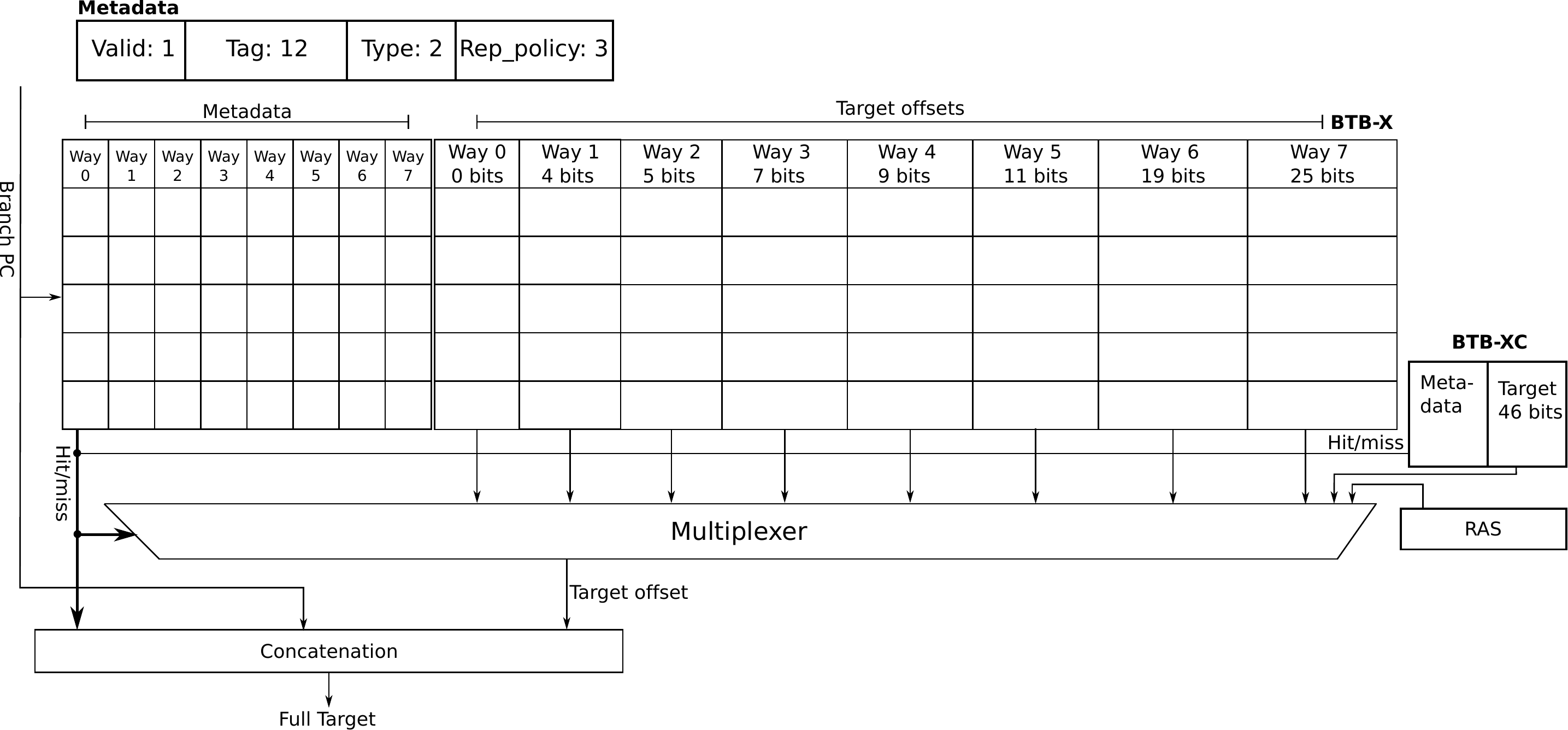}
    \caption{BTB-X organization and entry/set composition.} 
    \label{fig:btbx}
\end{figure*}


\runinsec{Associative searches} On allocating a new BTB entry, all BTB partitions (Main-BTB, Page-BTB, and Region-BTB) may need to be updated. The replacement policy chooses the entry to be replaced in Main-BTB. However, Page- and Region-BTB need to be searched to check if the page/region number for the incoming target is already present or not. As the page/region number can be present anywhere in the Page/Region-BTB, ITTAGE~\cite{ittage} uses fully-associative searches which increase the power requirements especially when the number of entries grows. PDede (partially) solves this limitation by restricting the number of entries where a page number can reside for a given branch target to 16. However, limiting the number of entries increases the likelihood of conflict misses.

\runinsec{Suboptimal storage utilization} Though R-BTB and PDede significantly improve storage utilization over conventional BTB, they still miss plenty of opportunity. This is because, in case of R-BTB, it uses a fixed size target representation for all branches, i.e., a 10-bit offset plus a fixed sized page-BTB pointer. In contrast, our analysis of Section~\ref{sec:analysis} shows a large variance in target offset sizes that naturally makes the single sized organization of R-BTB storage inefficient as it needs to be sized for the largest target offset. For example, as shown by Figure~\ref{fig:offsets}, 54\% of target offsets fit in six or fewer bits; however, R-BTB needs to use all 10+ bits for these branches, thus resulting in a high storage under-utilization. PDede provides slightly better storage utilization than R-BTB as it has differently sized entries for Same-page and Different-page branches. However, BTB entries of only two different sizes, i.e. Same-page and Different-page, are not enough to capture the large offset size variance (Figure~\ref{fig:offsets}) observed in server applications.

%% file: 5_design.tex
\section{BTB-X}
\label{sec:design}

BTB-X is a simple and storage-effective BTB organization. Building on the insights gained in Section~\ref{sec:analysis}, it stores target offsets, instead of full targets, to minimize storage requirements while also accounting for the large variance in target offset sizes. Its microarchitecture and entry/set composition are shown in Figure~\ref{fig:btbx}.

\subsection{BTB-X Organization}

The offset field in BTB-X needs to accommodate the uneven distribution of target offset sizes as observed in Section~\ref{sec:analysis}. In principle, this field should be sized such that the BTB can store the largest target offset. However, as the largest offset can be nearly as large as the full target, sizing the \textit{offset} field this way would nearly eliminate the potential storage saving from storing offsets. An attractive alternative is to size the offset field such that it can store the majority of offsets. Looking at Figure~\ref{fig:offsets}, an \textit{offset} field of 25-bits would capture more than 99\% of branch target offsets as they requires 25-bits or fewer. However, there are two major drawbacks to this scheme. First, it still leads to poor storage utilization. This is because, as Figure~\ref{fig:offsets} implies, 54\% of branches would waste more than three quarters of \textit{offset} storage capacity as they require only 6-bits or fewer for their offsets. Another 25\% of branches would waste nearly half of the offset storage as their offsets fit in 7-12 bits. Second, all branches that need more than 25 bits for their offsets can not fit in the BTB and will always cause BTB misses. Though, as there are very few such branches (< 1\%), their impact is likely to be small.

To minimize the storage under-utilization, we size different ways of a set associative BTB-X to hold different sized target offsets. A branch is allocated to a way whose offset field is at least as large as the number of bit required to store the target offset. We use an 8-way set associative BTB-X and leverage the data in Figure~\ref{fig:offsets} to appropriately size the offset field of each way such that each way covers about 12.5\% dynamically executed branches. Figure~\ref{fig:offsets} shows that, on average, 0-, 4-, 5-, 7-, 9-, 11-, 19-, and 25-bit offsets cover about 20\%, 36\%, 46\%, 61\%, 72\%, 79\%, 90\%, and 99\% dynamic branches. Therefore, we size the 8-ways of BTB-X ways to hold 0-, 4-, 5-, 7-, 9-, 11-, 19-, and 25-bit target offsets respectively. Notice that about 20\% of dynamic branches that require 0-bits for their offset are return instructions that read their target from RAS, as discussed in Section~\ref{sec:background}. Therefore, way-0 of BTB-X does not feature any storage for target offsets. Though return instruction do not get their target from BTB, they still need to be allocated BTB entries so that the branch prediction unit can identify them and pick their target from RAS while generating instruction stream to be fetched.

BTB-X covers 99\% of the dynamically executed branches and we employ a very small conventional direct-mapped BTB, called BTB-XC, that stores full target addresses for the remaining 1\% branches. Reserving a way in BTB-X for such branches would unnecessarily increase the storage requirements as these branches require much fewer entries than the number of sets in BTB-X. Indeed, based on our analysis, we size BTB-XC to store 64x fewer entries than BTB-X, i,e, 8x fewer entries than the number of sets in BTB-X.

\subsection{Accessing BTB-X}

\runinsec{BTB-X Lookup}
A BTB-X lookup is very similar to a conventional BTB lookup as shown in Figure~\ref{fig:btbx}. It is accessed with the index bits of a PC and all eight ways are looked up in parallel. Also, BTB-XC is looked up in parallel with BTB-X. The main difference between a conventional BTB lookup and BTB-X lookup is that a BTB-X lookup provides target offset, rather than full target address, if the lookup hits in way-1 to way-7. Thus, target offset needs to be concatenated with branch PC to get the full target address. The number of bits to be concatenated from branch PC depends on the BTB-X way in which the lookup hits. For example, a hit in way-1 provides 4 lower order bits of target while the rest needs to be concatenated from branch PC. Further, a hit in way-0 implies that the full target is in RAS, while a hit in BTB-XC provides the full target address.

\vspace{0.05in}

\runinsec{BTB-X Allocation}
As with any existing BTB organization, BTB-X entries are allocated (or updated) as branch instructions retire. The number of bits required to represent a branch target offset determines the way(s) where a branch can be allocated an entry. For example, return instructions can be allocated entries in any of the ways, based on replacement policy's decision, as they have no offset and can fit in all ways. Other branches have fewer ways where they can be allocated entries that are determined by the minimum number of bits required to store their offsets. For example, if a branch requires 20 bits for its target offset, it cannot be allocated in way-0 to way-6.

BTB-X uses a slightly modified least recently used (LRU) replacement policy. Concretely, we modify it to compare the LRU counters of only the entries that can accommodate the target offset and replace the one that is least recently used among them. All other aspect of LRU, such as counter updates, stay exactly the same as in baseline policy.

%% file: 6_eval.tex
\section{Evaluation}
\label{sec:eval}

\begin{small}
\begin{table}[t]
	\centering 
	\caption{Microarchitectural parameters}
	{\small
		\scalebox{1} {		
		\begin{tabular}{|c|c|}
			\hline  
			\hline  

            \textbf{Parameter} & \textbf{Value} \\
			\hline
            Fetch & 6-wide, 128-instruction FTQ \\
			\hline  			
            Branch Predictor& Hashed Perceptron \\
			\hline
			Return address stack& 64 entries \\
			\hline
            Scheduler& 128 entries \\
			\hline
            Re-order buffer& 352 entries \\
			\hline
			Load queue& 128 entries \\
			\hline
			Store queue& 72 entries \\
			\hline

			L1-I & 
            \begin{tabular}{@{}c@{}}
            32 KB, 8-way, \\4 cycle latency, 8 MSHRs
            \end{tabular} \\
			\hline
			L1-D & 
            \begin{tabular}{@{}c@{}}
            48 KB, 12-way, \\5 cycle latency, 16 MSHRs
            \end{tabular} \\
            \hline
			L2 & 
            \begin{tabular}{@{}c@{}}
            512 KB, 8-way, \\14/15 cycle latency, 32 MSHRs
            \end{tabular} \\
			\hline
			LLC & 
			\begin{tabular}{@{}c@{}}
			2MB, 16-way, \\34/35 cycle latency, 64 MSHRs
			\end{tabular}\\
			\hline
			\hline  			
		\end{tabular}
	  }
	}
	\label{table:microParam}

\end{table}
\end{small}

\subsection{Methodology}
We use ChampSim~\cite{champsim} to evaluate the efficacy of BTB-X on server and client workload traces provided by Qualcomm for the first Instruction Prefetching Championship (IPC-1)~\cite{ipc1}. We warm up microarchitectural structures for 50M instructions and collect statistics over the next 50M. The microarchitectural parameters for the modeled processor, resembling Intel Sunny Cove~\cite{sunnycove}, are listed in Table~\ref{table:microParam}.

We improved two important aspects of Champsim to evaluate the baseline, state-of-the-art, and proposed BTB organizations. First, being a trace-driven simulator, Champsim detects branches by consulting the information available in the traces, rather than looking up a BTB. This essentially translates to Champsim using an ideal BTB. Therefore, we first implement a realistic conventional BTB (Conv-BTB), presented in Section~\ref{sec:background}, in Champsim. Second, Champsim resolves all branches in execute stage, i.e., branch mispredictions are detected and the fetch is resteered to correct path only when a mispredicted branch instruction reaches the execute stage. Such branch resolution overestimates the misprediction penalty of unconditional direct branches. This is because such branches can be resolved in the decode stage (hence, fetch can be resteered sooner) as they are always taken, thus the PC of the next instruction can be compared to the target encoded in the branch instruction to detect mispredictions. Further, taken conditional branches that miss in BTB but are correctly predicted by the direction predictor, can also be resolved in the decode stage. To do so, the fetch stage passes the direction prediction for all instructions, despite BTB hit/miss, to decode stage. If decode identifies a branch that missed in the BTB but predicted taken by the direction predictor, it resteers the fetch to the target encoded in the branch instruction, thus reducing BTB miss penalty. Given that the direction predictors are highly accurate, this optimization reduces average BTB miss penalty. Overall, we improve branch resolution so that the unconditional direct branches and the taken conditional branches that miss in BTB are resolved in the decode stage. Finally, BTB is updated at commit stage by only the taken branches (both conditional and unconditional).

\begin{small}
\begin{table}[t!]
  \centering
  \caption{BTB-X storage requirements. The numbers in parentheses are for BTB-XC.}
  \label{table:metadata}
  \begin{tabular}{lrrr} \hline
    \textbf{Entries} & \textbf{Sets} & \textbf{Set size}
    & \textbf{Storage} \\\hline
    256(4) & 32(4) & 224(64)-bits & 0.9KB\\\hline
    512(8) & 64(8) & 224(64)-bits & 1.8KB\\\hline
    1K(16) & 128(16) & 224(64)-bits & 3.6KB\\\hline
    2K(32) & 256(32) & 224(64)-bits & 7.25KB\\\hline
    4K(64) & 512(64) & 224(64)-bits & 14.5KB\\\hline
    8K(128) & 1024(128) & 224(64)-bits & 29KB\\\hline
    16K(256) & 2048(256) & 224(64)-bits & 58KB\\\hline
  \end{tabular}
  \vspace{-0.1in}
\end{table}
\end{small}

\begin{small}
\begin{table*}
  \centering
  \caption{Number of branches in different BTB organizations at various storage budgets.}
  \label{table:btbComp}
  \begin{tabular}{cccccc} \hline
    \textbf{Storage} & \textbf{BTB-X + BTB-XC} & \textbf{PDede} & \textbf{Conv-BTB}\\\hline
    \begin{tabular}{r} 
    \textbf{}\\\hline
    0.9KB\\
    1.8KB\\
    3.6KB\\
    7.25KB\\
    14.5KB\\
    29KB\\
    58KB
    \end{tabular}
    
    &
    \begin{tabular}{r} 
    \textbf{Branches}\\\hline
    256 + 4\\
    512 + 8\\
    1K + 16\\
    2K + 32\\
    4K + 64\\
    8K + 128\\
    16K + 256
    \end{tabular}
    
    &
    \begin{tabular}{rrrr} 
    Page-BTB budget & Main-BTB budget & Entry Size&\textbf{Branches}\\\hline
    0.078KB&0.817KB&32-bits&210\\
    0.156KB&1.645KB&32.5-bits&415\\
    0.312KB&3.3KB&33-bits&820\\
    0.625KB&6.6KB&33.5-bits&1617\\
    1.25KB&13.2KB&34-bits&3190\\
    2.5KB&26.5KB&34.5-bits&6292\\
    5KB&53KB&35-bits&12405\\
    \end{tabular}
    
    &
    \begin{tabular}{rr} 
    Entry Size&\textbf{Branches}\\\hline
    64-bits&116\\
    64-bits&232\\
    64-bits&464\\
    64-bits&928\\
    64-bits&1856\\
    64-bits&3712\\
    64-bits&7424
    \end{tabular}\\\hline
    
  \end{tabular}
  \vspace{-0.2in}
\end{table*}
\end{small}

\subsection{Storage breakdown}
\label{sec:storageBreak}

We first assess the number of branches different BTB organizations (Conv-BTB, PDede, and BTB-X) can accommodate in a given storage budget compared to each other. We use storage budgets required for storing 256, 512, 1K, 2K, 4K, 8K, and 16K branches in BTB-X as presented in \Cref{table:metadata}. Our calculations assume a 48-bit virtual address space and BTB-X entry compositions presented in Figure~\ref{fig:btbx}. To double the number of entries in BTB-X, we double the number of sets while keeping the associativity same. Notice that \Cref{table:metadata} presents set size instead of entry size. This is because BTB-X features different sized entries in different ways; however, the set size remains constant.
 
\Cref{table:btbComp} presents the number of branches the different BTB organizations can track at different storage budgets. PDede distributes the overall BTB storage budget among its Main-BTB, Page-BTB, and Region-BTB. We follow the distribution used by its inventors\cite{pdede} to allocate the budget among different PDede BTBs as shown in \Cref{table:btbComp}. Accordingly, for 29KB storage budget, we configure PDede to use 1K Page-BTB entries and about 6K Main-BTB entries. While halving the storage budget to lower values, we halve the number of entries in the Main-BTB as well as the Page-BTB. Halving the number of Page-BTB entries reduces the number of bits required to store Page-BTB pointer in the Main-BTB. Thus, the Main-BTB entry size reduces with the reduction in storage budget. Further, we use four Region-BTB entries across all storage budgets, so Region-BTB requires a fixed storage of 0.0107KB. Also recall that PDede reserves half of the ways in a set for same-page branches while the other half can store both same-page and different-page branches. Therefore, its entries are of two different sizes. The PDede entry size shown in \Cref{table:btbComp} is the average of two sizes.

As the table shows BTB-X stores significantly more branches than any other BTB organizations. Concretely, it stores 2.24x more branches than a conventional BTB organization. Compared to PDede, BTB-X stores 1.24x more branches at 0.9KB storage budget and 1.34x more branches at 58KB storage budget. BTB-X's advantage over PDede increases with storage budget because PDede entries require more bits at higher budgets to accommodate larger Page-BTB pointers. 

\begin{figure*}[t!]
    \centering
    \includegraphics[width=0.9\textwidth, trim=60 300 50 340, clip]{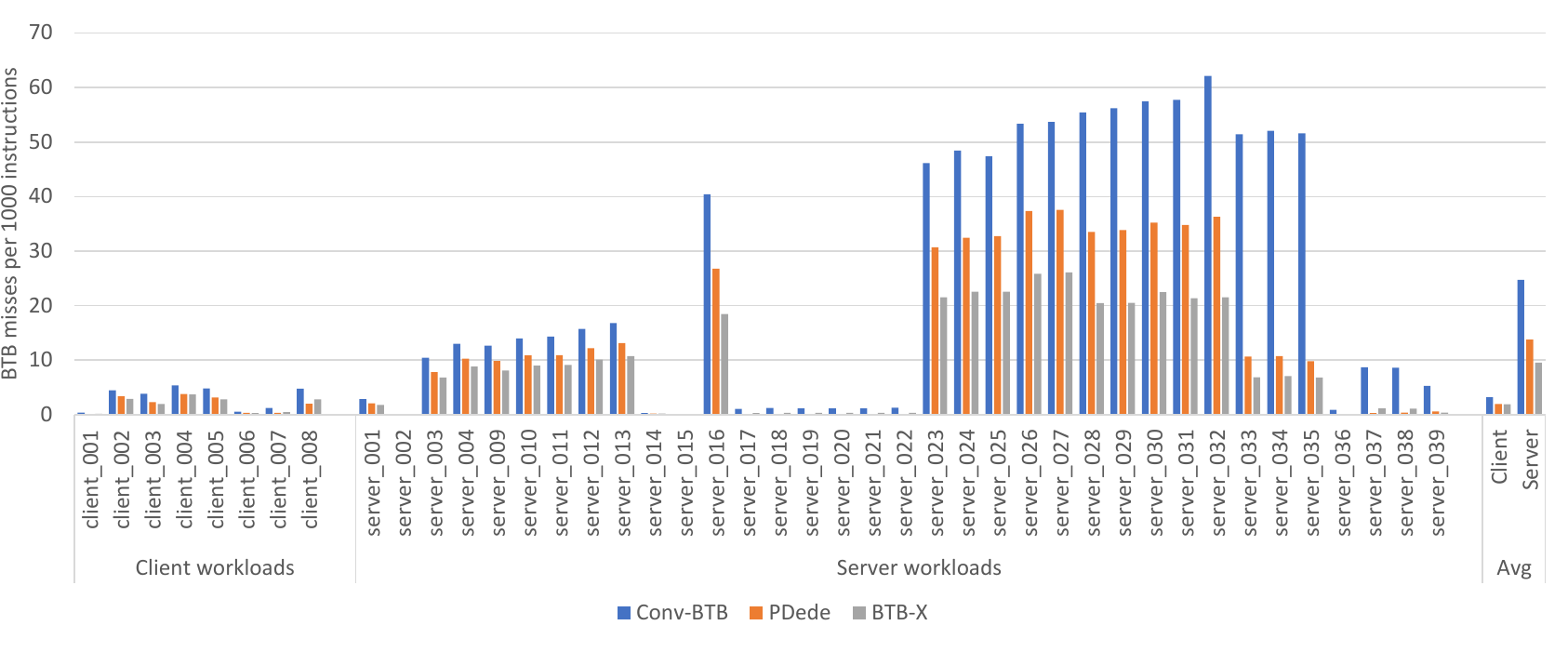}
    \vspace{-0.1in}
    \caption{BTB MPKI experienced by different BTB organizations.}
    \vspace{-0.2in}
    \label{fig:mpki}
\end{figure*}

\begin{figure*}[t!]
    \centering
    \includegraphics[width=0.9\textwidth, trim=60 300 50 340, clip]{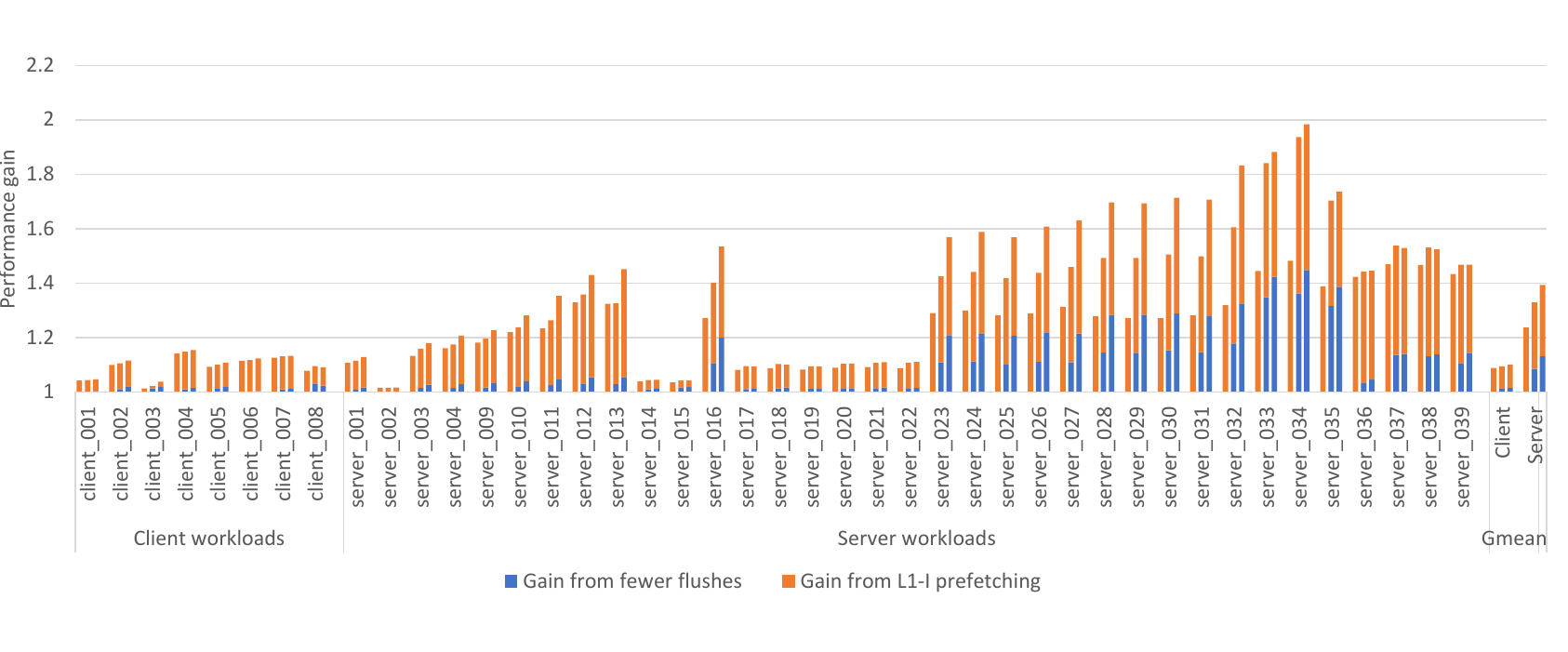}
    \vspace{-0.1in}
    \caption{Performance gain obtained by conventional BTB (with FDIP), PDede and BTB-X (with and without FDIP) over the conventional BTB without FDIP. The three bars for each workload correspond to Conv-BTB, PDede, and BTB-X respectively.}
    \label{fig:perfAll}
\end{figure*}

\subsection{BTB MPKI}
\label{sec:mpki}

To understand the advantage of higher BTB-X branch density, we measure misses per 1000 instructions (MPKI) that different BTB organizations incur on client and server workloads. Since BTB misses for not-taken branches do not hurt performance, we only consider the BTB misses for taken branches. For this analysis, we assume a BTB storage budget of 14.5KB that corresponds to 4160-, 3190-, and 1856-entries in BTB-X, PDede and Conv-BTB respectively. The results are presented in Figure~\ref{fig:mpki}. 

As the figure shows, server workloads experience significantly higher MPKI compared to client workload due their massive instruction and branch footprints. The figure also shows that BTB-X provides a much lower MPKI compared to both conventional BTB and PDede especially on server workload. Concretely, on average, conventional BTB incurs 25 MPKI on server workload as it stores the least amount of branches among the three organization for a given storage budget. PDede is able to lower the MPKI to 13.7 while BTB-X brings it further down to 9.5. The advantage of BTB-X over other organizations is particularly evident on very high MPKI workloads, i.e., server\_023 to server\_035, where it provides much lower MPKI compared to conventional BTB and PDede.

\subsection{Performance}
\label{sec:perfAll}

To assess how the reduced MPKI translates to performance, we compare the performance of the three BTB organizations on client and server workloads. Recall from Section~\ref{sec:background} that a larger BTB delivers two distinct benefits: 1) it reduces the incidence of pipeline flushes by detecting branches in the upcoming control flow and 2) it facilitates instruction prefetching when coupled with FDIP. Thus, we compare the performance gains achieved by the three BTB organizations by evaluating them with FDIP. 

Figure~\ref{fig:perfAll} presents the performance gains obtained on server and client traces. The results are normalized to the performance of the Conv-BTB without any instruction prefetching. The figure shows three bars for each workload. The first bar presents performance gain achieved by Conv-BTB when copuled with FDIP. The second and third bars present the performance gains achieved by PDede and BTB-X respectively. The PDede and BTB-X bars divide the performance gain into contributions from fewer pipeline flushes and from better instruction prefetching stemming from capturing more branches in the BTB.

Looking at the overall performance gain with instruction prefetcher (FDIP), the figure shows that BTB-X provides a geometric mean gain of 39\% over baseline on server workloads. In comparison, PDede and Conv-BTB deliver a performance gain of only 33\% and 24\% on these workloads. Looking at individual workloads, BTB-X comprehensively outperforms PDede and Conv-BTB on server\_023 to sever\_32. For example, on server\_032, BTB-X provides 83\% speedup over baseline whereas PDede and Conv-BTB achieve only 60\% and 32\% performance gain. This is because the branch working set of these workloads starts to fit in BTB-X due to its higher branch capacity. As a result, BTB MPKI lowers which not only reduces pipeline flushes and but also keeps FDIP on correct prefetch path for longer intervals. 

Looking at the results without instruction prefetcher, Figure~\ref{fig:perfAll} shows that BTB-X provides 13\% performance gain over the baseline Conv-BTB whereas PDede is achieves 8\% gain. On individual workloads, BTB-X achieves significantly high gain over Conv-BTB and PDede on workloads from server\_23 to server\_32 even without FDIP. Figure~\ref{fig:perfAll} also shows the FDIP by itself performs better with more number of BTB entries. For example, on server\_32 FDIP with Conv-BTB provides 32\% performance gain. With PDede, the performance gain from prefetching increases to 42\% and with BTB-X it further increases to 51\%.

\begin{small}
\begin{table}[t!]
\centering
\caption{Energy requirements of different BTB designs.}
\label{tab:energy}
        \begin{tabular}{llrrr} \hline
            \textbf{BTB} & \textbf{Access Type} & \textbf{Energy} & \textbf{\#Accesses} & \textbf{Energy}\\
             &  & \textbf{(Per access)} &  & \textbf{(Total)}\\\hline
            \multirow{3}{*}{Conv-BTB} & \multicolumn{1}{l}{Read} & \multicolumn{1}{r}{13.2pJ} & \multicolumn{1}{r}{1.60E+08} & \multicolumn{1}{r}{2122µJ}\\\cline{2-5}
                                      & \multicolumn{1}{l}{Write} & \multicolumn{1}{r}{25.2pJ} & \multicolumn{1}{r}{4.36E+06} & \multicolumn{1}{r}{110µJ}\\\cline{2-5}
                                      & \multicolumn{1}{l}{\textbf{Total Energy}} & \multicolumn{1}{r}{} & \multicolumn{1}{r}{} & \multicolumn{1}{r}{\textbf{2232µJ}}\\\hline

            \multirow{6}{*}{PDede} & \multicolumn{1}{l}{Main-BTB Read} & \multicolumn{1}{r}{8.4pJ} & \multicolumn{1}{r}{1.24E+08} & \multicolumn{1}{r}{1047µJ}\\\cline{2-5}
                                 & \multicolumn{1}{l}{Main-BTB Write} & \multicolumn{1}{r}{12.5pJ} & \multicolumn{1}{r}{5.74E+05} & \multicolumn{1}{r}{7µJ}\\\cline{2-5}
                                 & \multicolumn{1}{l}{Page-BTB Read} & \multicolumn{1}{r}{0.9pJ} & \multicolumn{1}{r}{2.01E+06} & \multicolumn{1}{r}{2µJ}\\\cline{2-5}
                                 & \multicolumn{1}{l}{Page-BTB Write} & \multicolumn{1}{r}{0.8pJ} & \multicolumn{1}{r}{2.04E+04} & \multicolumn{1}{r}{0.02µJ}\\\cline{2-5}
                                 & \multicolumn{1}{l}{Page-BTB Search} & \multicolumn{1}{r}{6.2pJ} & \multicolumn{1}{r}{2.14E+05} & \multicolumn{1}{r}{2µJ}\\\cline{2-5}
                                 & \multicolumn{1}{l}{\textbf{Total Energy}} & \multicolumn{1}{r}{} & \multicolumn{1}{r}{} & \multicolumn{1}{r}{\textbf{1058µJ}}\\\hline
                                 
            \multirow{3}{*}{BTB-X} & \multicolumn{1}{l}{Read} & \multicolumn{1}{r}{8.5pJ} & \multicolumn{1}{r}{1.16E+08} & \multicolumn{1}{r}{994µJ}\\\cline{2-5}
                                 & \multicolumn{1}{l}{Write} & \multicolumn{1}{r}{11.4pJ} & \multicolumn{1}{r}{4.03E+05} & \multicolumn{1}{r}{5µJ}\\\cline{2-5}
                                 & \multicolumn{1}{l}{\textbf{Total Energy}} & \multicolumn{1}{r}{} & \multicolumn{1}{r}{} & \multicolumn{1}{r}{\textbf{999µJ}}\\\hline

        \end{tabular}
\end{table}
\end{small}

These results show that by accommodating more branches in a given storage budget, BTB-X not only reduces pipeline flushes but also improves instruction prefetching, both lead to better performance.

\begin{figure*}
    \centering
    \begin{subfigure}[t]{0.935\columnwidth}
        \centering
        \includegraphics[width=0.935\columnwidth, trim=70 235 60 250, clip]{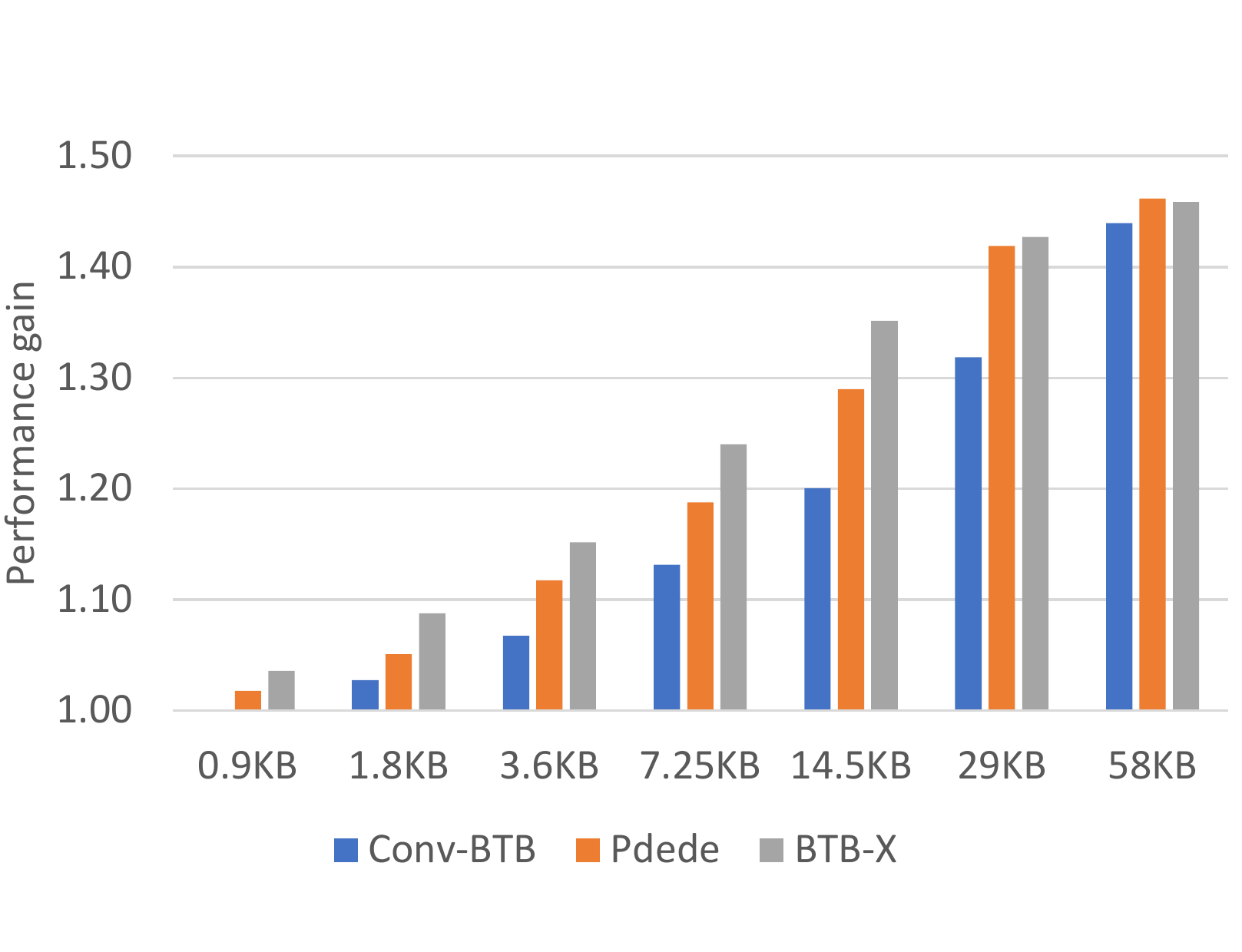}
        \caption{Server workloads}
        \label{fig:serverPerf}
    \end{subfigure}
    ~ 
    \begin{subfigure}[t]{0.935\columnwidth}
        \centering
        \includegraphics[width=0.935\columnwidth, trim=70 235 60 250, clip]{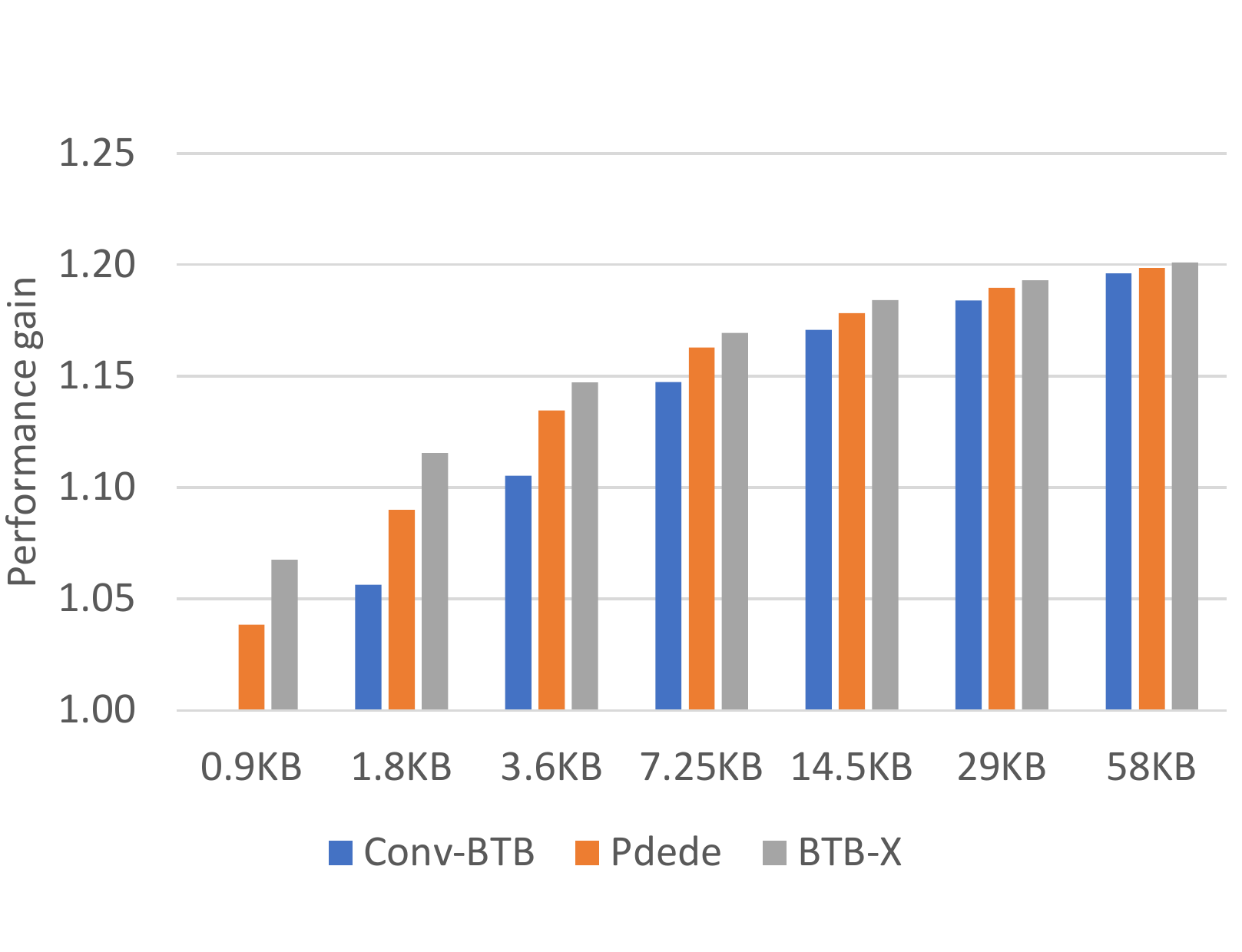}
        \caption{Client workloads}
        \label{fig:clientPerf}
    \end{subfigure}
    \caption{Performance gains for conventional BTB, PDede, and BTB-X on (a) \textbf{server} and (b) \textbf{client} workloads over a conventional BTB with 0.9KB storage budget. X-axis label is storage requirements of 256-, 512-, 1K-, 2K-, 4K-, 8K-, and 16K-entry BTB-X.}\label{fig:perf}
\end{figure*}

Finally, Figure~\ref{fig:perfAll} shows that all three BTB organizations perform similar on client workloads. This is because their branch working sets mostly fit in the baseline Conv-BTB and the additional entries in PDede and BTB-X do not bring much performance benefit.

\subsection{Energy and delay analysis}

We use Cacti 7.0~\cite{cacti} to analyze the energy requirements and access latencies of Conv-BTB, PDede, and BTB-X at 22 nm, which is the most recent technology node supported by Cacti. For this analysis we assume the same storage budget, i.e. 14.5KB, as used in Section~\ref{sec:perfAll} for the performance analysis.

\vspace{0.05in}

\runinsec{Energy requirements}Table~\ref{tab:energy} shows the per access read and write energy requirements of different BTB designs. As the table shows, BTB-X and PDede's Main-BTB incur very similar per access read and write energy cost. However, in addition to Main-BTB, PDede also needs to access Page-BTB for different-page branches, i.e., the branches that have their targets in a different page than the branches themselves. Further, the Page-BTB needs to be searched on a BTB write to check if the target page number is already in Page-BTB or not. Consequently, PDede's per access read and write energy for different page branches reaches 9.3 pico Joules (pJ) and 19.5 pJ, respectively, compared to 8.5pJ and 11.4 pJ of BTB-X. PDede also features a Region-BTB; however, its energy requirements are negligible and, thus, not shown in Table~\ref{tab:energy}. Finally, Conv-BTB's per access energy cost is significantly higher than BTB-X as its each read and write access requires 13.2pJ and 25.2pJ respectively.

Table~\ref{tab:energy} also shows the number of read/write accesses, averaged across the workloads, and the total energy consumption. Despite very similar per access energy cost, PDede' Main-BTB consumes considerably higher energy than BTB-X. This is because PDede often goes on the wrong execution path due to its higher MPKI. These additional wrong path BTB accesses, reflected in higher BTB reads in Table~\ref{tab:energy}, result in higher energy consumption. Further, PDede needs to handle more BTB writes than BTB-X because it holds fewer branches, which results in frequent replacements. Thus, the total energy consumption of PDede reaches 1058µJ compared to 999µJ of BTB-X. Finally, the energy requirements of Conv-BTB are significantly higher, 2232µJ, than BTB-X because of higher per access energy and higher number of total accesses.

Overall, this analysis shows that BTB-X not only delivers better performance than PDede but also consumes less energy, thus providing much better energy efficiency.

\vspace{0.05in}

\runinsec{Access Latency}Our analysis shows that the Conv-BTB requires about 0.36ns to complete an access. As discussed in Section~\ref{sec:pdedearch}, PDede's access latency is the sum of Main-BTB and Page-BTB access latencies as these two structures are accessed sequentially. Our analysis shows that the Main-BTB and Page-BTB accesses require 0.34ns and 0.13ns, respectively, thus resulting in an overall PDede access latency of 0.47ns which is considerably higher than Conv-BTB latency. To address this, PDede employs multi-cycle BTB accesses: the Main-BTB is accessed in the first cycle, and the Page-BTB is accessed in the next cycle only if the branch is predicted to be taken and it's target is in a different page than the branch. Thus, the same page branches need one cycle and the taken different page branches need two cycles to get their target address from PDede. Finally, our analysis shows that a BTB-X access takes only 0.33ns. In summary, this analysis shows that BTB-X provides better storage efficiency without any adverse effects on the access latency.

\subsection{Performance variation with BTB storage budget}

To further understand the performance advantage of BTB-X over PDede and Conv-BTB, we compare their performances across different storage budgets. Figure~\ref{fig:perf} presents the performance gains obtained on server and client workloads. The results are normalized to the performance of Conv-BTB with 0.9KB storage budget. Instruction prefetching is enabled in all designs including baseline.

As the figure shows, on server workloads, BTB-X provides significantly higher performance than the Conv-BTB and PDede for equal storage budgets of up to 29KB and 14.5KB respectively. The performance advantage of BTB-X is pronounced on server traces whose large instruction footprints pressure the BTB and L1-I. For instance, BTB-X provides 35\% performance gain over the baseline compared to 29\% and 20\% of PDede and Conv-BTB respectively at 14.5KB budget. At large BTB storage budgets, the branch working sets of many workloads start to fit in the available BTB capacity, at which point the performance gap between BTB-X and the other two designs diminishes. Also, the performance gap between the three BTB organizations levels off earlier on client trace due to their smaller instruction working sets.

A key take-away from this figure is that BTB-X outperforms the conventional BTB even when it is given just half the storage budget of its conventional counterpart. For example, in Figure~\ref{fig:serverPerf}, the Conv-BTB improves performance by 20\% with a 14.5KB budget whereas BTB-X provides a 24\% improvement with just 7.25KB. The reason for this phenomenon is that BTB-X accommodates 2.24x more entries than Conv-BTB of equal storage budget; thus, halving BTB-X's budget still gives a slight capacity advantage over Conv-BTB. 

\subsection{Analyzing target offset distribution in more workloads}

\begin{figure}
\centering
\includegraphics[width=\columnwidth, trim=70 90 60 100, clip]{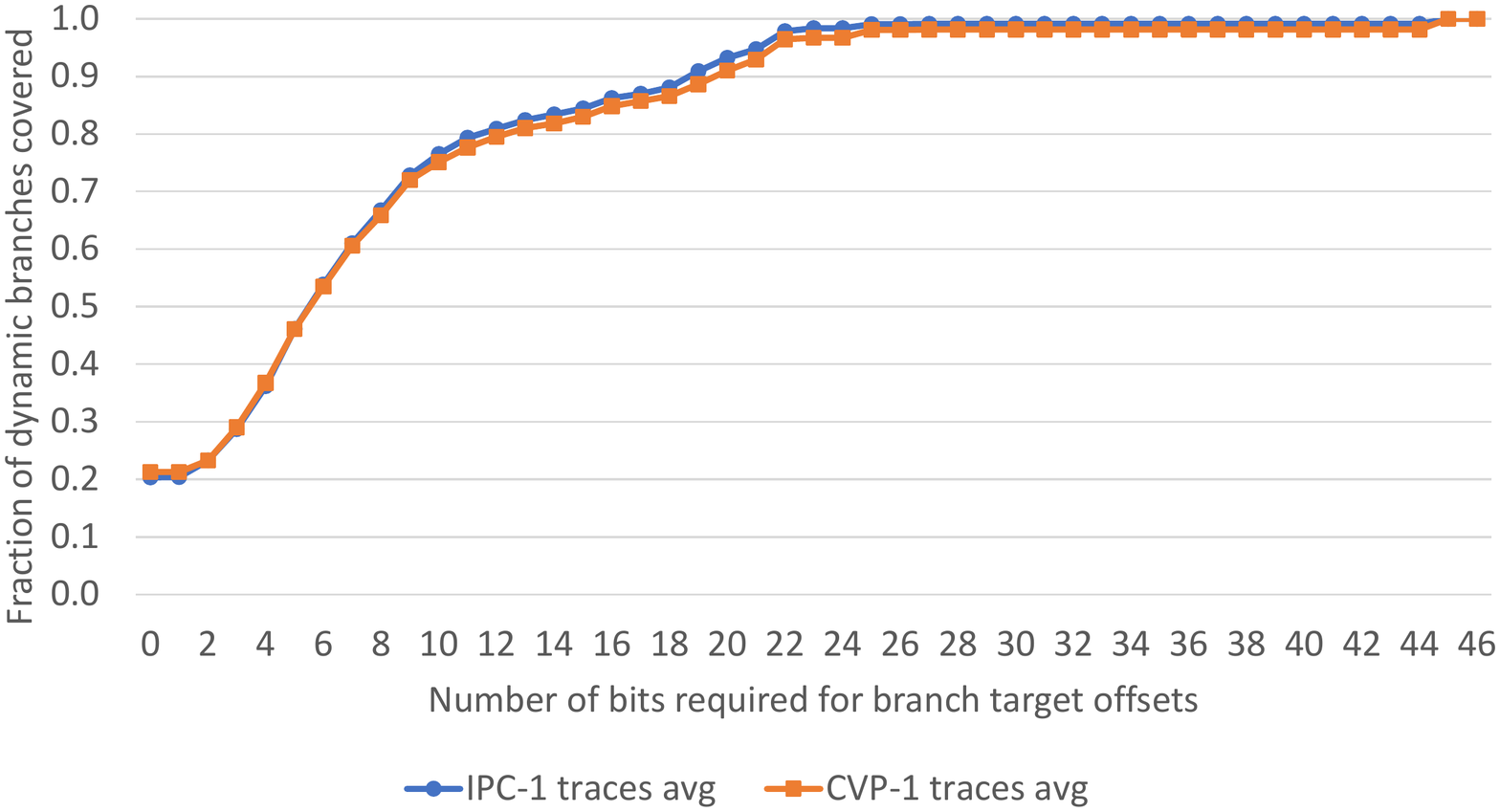}
\caption{Target offset distribution in CVP-1 and IPC traces.}
\label{fig:cvp_res}
\end{figure}

We study the target offset distribution in 750+ Qualcomm server traces that were provided for the first Championship Value Prediction(CVP-1)~\cite{cvp}. The results, presented in Figure~\ref{fig:cvp_res}, show that their offset distribution is very similar to the distribution in IPC-1 traces presented in Figure~\ref{fig:offsets}. This study confirms that such an offset distribution is a consequence of how the applications are written and the resulting control-flow behavior. As discussed in Section~\ref{sec:analysis}, such offset distribution stems from the fact that the conditional branches dominate dynamic branch working set and they tend to have short offsets. This is because conditional branches guide the control-flow inside functions, and software engineering principles favor small functions. Consequently, short offsets dominate the branch offset distribution.

\begin{figure}
\centering
\includegraphics[width=\columnwidth, trim=70 90 60 100, clip]{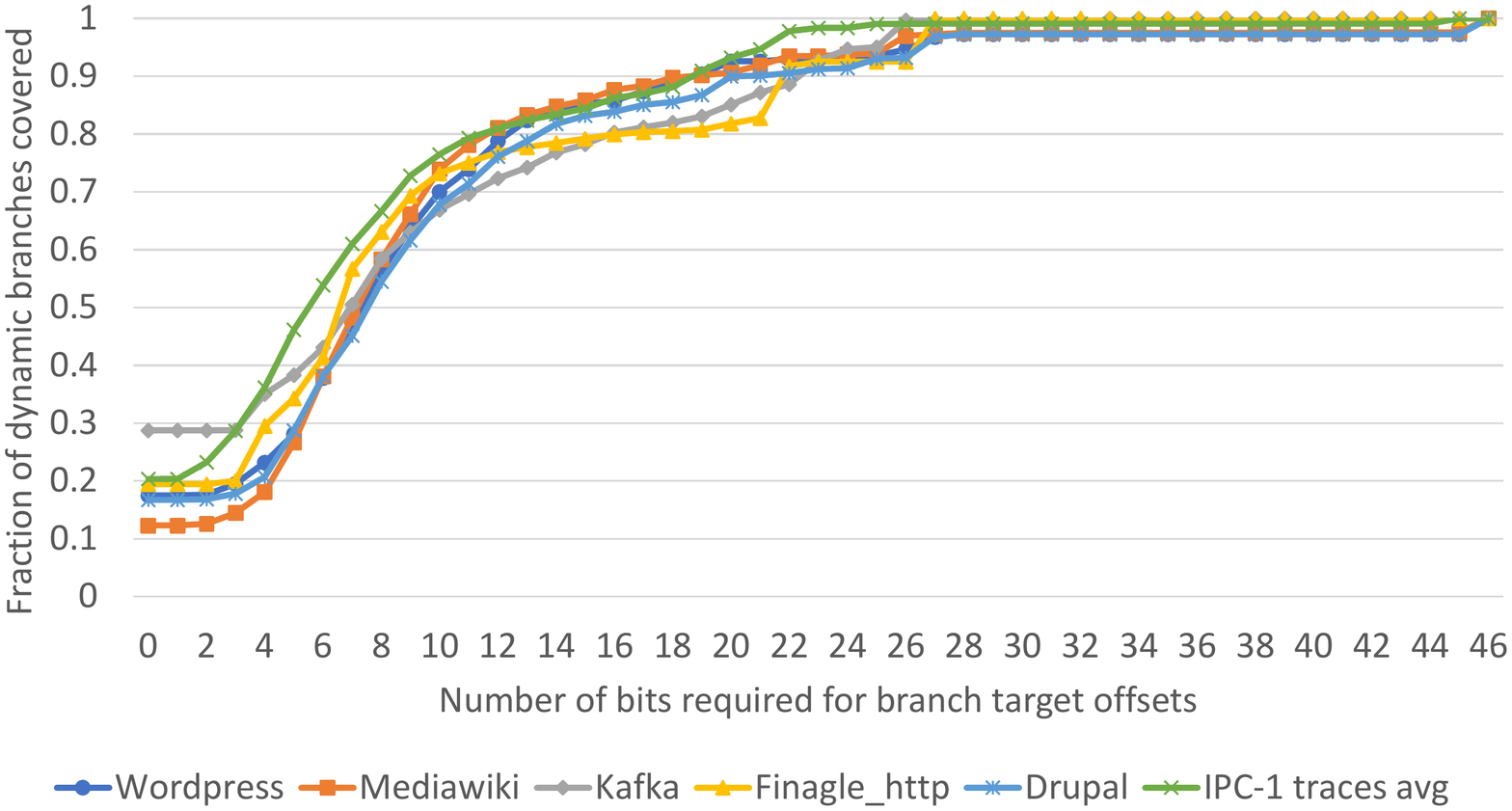}
\caption{Target offset distribution in x86 compiled server applications and Arm64 IPC traces.}
\label{fig:x86_res}
\end{figure}

In addition to CVP-1 traces, we analyze five more server applications - Wordpress~\cite{wiki:wordpress}, Mediawiki~\cite{wiki:mediawiki}, and Drupal~\cite{wiki:drupal} from Facebook’s HHVM OSS-performance benchmarks~\cite{fboss}, Kafka~\cite{wiki:kafka} from Java DaCapo~\cite{blackburn2006dacapo}, and Finagle-HTTP~\cite{finagle-http} from Java Renaissance~\cite{Prokopec2019}. Further, these applications are compiled to x86 (CVP-1 and IPC-1 traces are compiled to Arm64) which also enables us to assess the impact of ISA on target offset distribution. The results presented in Figure~\ref{fig:x86_res} show that the offset distribution in these applications is also very similar to that in IPC-1 traces. The only difference is that x86 traces require slightly larger offsets (1 or 2-bits more) to achieve a similar dynamic branch coverage as the Arm64 (CVP-1 and IPC-1) traces. For example, 6-bit offsets cover about 54\% branches in Arm64 traces, whereas x86 offsets need 8-bits to achieve 58\% branch coverage. This is because x86 offsets specify the distance between branch PC and target in number of \emph{bytes} because x86 instruction are variable size. In contrast, Arm64 offsets specify this distance in number of \emph{instructions} because all instructions are 4-bytes, thus saving 2 offset bits.

As BTB-X needs to store slightly larger offsets for x86 than Arm64, we reassess its storage advantage over PDede and Conv-BTB for x86 architectures. As each way in 8-way BTB-X needs to cover about 12.5\% of branches, we size its ways to store offset of 0-, 5-, 6-, 7-, 9-, 12-, 20-, and 27-bits based on the offset distribution in x86 applications shown in Figure~\ref{fig:x86_res}. Thus, each set needs 86-bits for offsets compared to 80-bit in Arm64. Consequently, BTB-X's storage advantage is slightly lower for x86 than Arm64. However, BTB-X still stores 2.18x more branches than Conv-BTB for x86 (2.24x for Arm). Compared to PDede, BTB-X stores 1.21x more branches (1.24x for Arm64) at 0.9KB storage budget and 1.31x more branches (1.34x for Arm64) at 58KB storage budget. (Section~\ref{sec:storageBreak} presents this analysis of Arm64 traces.)

%% file: 7_related.tex
\section{Related work}
\label{sec:related}

\runinsec{Mitigating BTB misses}
BTB was first disclosed by Losq~\cite{losq} and was further expanded by Lee et al~\cite{btb}. Since BTB lies on the critical path for instruction delivery, there has been several proposals to increase its effectiveness. Instead of accessing BTB with the PC of each individual instruction, Yeh et al. \cite{bbtb} proposed to access it with basic-block address and store not only the target but also the fall-through address in the BTB. In case the branch is predicted to be not taken, the fall-through address is used, after fetching the current basic-block, as the next PC for both instruction fetch as well as for the next BTB access. The advantage of such a BTB organization over the conventional BTB organization is that it reduces BTB bandwidth and power requirements as a single access provides the next control flow divergence point, whereas the conventional organization requires as many accesses as the number of instructions until the next branch. Whereas the initial proposal on basic-block-based BTB~\cite{bbtb} stores full fall-through address, the later work~\cite{fbtb} proposed to store the delta between two sequential basic-block addresses. Fagin \cite{fagin1997partial} proposed to use the BTB storage more effectively by storing only the partial tags. To further amortize the tag storage cost, some designs proposed to share a BTB entry among multiple branches that reside in the same cache block \cite{amdbtb, confluence}. Though these BTB designs aim to improve different aspects of BTB management, they all share a common trait, i.e., they store full target addresses. Thus, the key idea of BTB-X can be applied to all of these BTB designs to reduce their target storage cost.

Prior work \cite{DUPN, ittage, jan, pdede, btbxCAL, btbxPACT} has also explored mechanisms to reduce the storage cost of branch targets. Seznec \cite{DUPN, ittage} proposed to break the target address into page number and offset; and store a pointer to the page number, along with the page offset, in the BTB while the page number itself is stored in a separate structure. It reduces the storage cost as a pointer to page number is smaller than the page number itself, and the page number for all the targets in a page is stored only once. Hoogerbrugge \cite{jan} proposed to size some of the entries in a set for storing small target offsets, thus reducing BTB storage requirements. 

The state-of-the-art BTB design, PDede \cite{pdede}, combines these two ideas to address their individual limitations. Concretely, Sezenc's design is sub-optimal for same-page branches as it unnecessarily stores (pointer to) their target page number even though it is same as the page number of their branch PCs. In contrast, Hoogerbrugge's design is sub-optimal for inter-page branches as it stores their full targets. Inspired from Hoogerbrugge's design, PDede sizes some entries in a set to store same-page branch targets; and similar to Seznec's design, for inter-page branches, it stores pointers to page numbers instead of page numbers themselves. PDede further reduces the inter-page target storage cost by dividing the page number into page- and region-number. However, as it is based on Seznec's design, it also has to pay the addition latency cost of indirection between main-BTB and the page-/region-BTB. Micro BTB~\cite{microbtb}, proposes a flexible BTB entry structure where each entry can store either one branch, if its offset is large, or two branches if their offsets are small. We show that all these designs are sub-optimal in exploiting the storage optimization opportunity presented by the uneven branch offset distribution. BTB-X not only captures this opportunity but also avoids the BTB indirection of the state-of-the-art.

Apart from optimizing BTB organization, prior work~\cite{phantom, ibmBTB, twig, boomerang, shotgun} has also explored BTB prefilling/prefetching to mitigate BTB misses. The state-of-the-art in BTB prefetching is a profile guided software prefetcher, called Twig~\cite{twig}. It analyzes an application's execution profile to identify critical BTB misses and then injects software prefetch instructions. The prefetch instruction takes compressed branch PC and target as operands and its execution fills this information in BTB. These prefetching techniques are complementary to BTB organization and, thus, can be used along with BTB-X.

\runinsec{Mitigating L1-I misses}
As L1-I misses continue to be a major performance limiter in server applications\cite{profileWarehouse, jukebox, wosc}, prior work has proposed both hardware and software mechanisms to mitigate L1-I misses. On the hardware side, state-of-the-art temporal stream prefetchers \cite{tifs, pif} record the L1-I miss/access history and replay it to discover prefetch candidates. While such prefetchers are highly effective, their huge metadata storage cost renders them impractical despite recent attempts to address this weakness \cite{shift, confluence}. Fetch-directed prefetchers use in-core structures (BTB and branch direction predictor) to run ahead of the fetch unit to find prefetch candidates. While the early work \cite{fdip} focused on L1-I prefetching only, the state-of-the-art fetch-directed prefetchers \cite{boomerang, shotgun} also prefill into the BTB.

Several purely-software based approaches to instruction prefetching and improving the L1-I capacity has also been proposed \cite{chen2016autofdo, li2010lightweight, ottoni2017optimizing, luk2004ispike, panchenko2019bolt, luk1998cooperative, annavaram2003call, ayers2019asmdb, khan2020spy}. These methods use data from application profiling to perform either compile-time, link-time or post-link time optimizations. Since these methods are software-only they will benefit from the increased BTB capacity provided by the BTB-X organization. 

%% file: 8_conclusion.tex
\section{Conclusion}
\label{sec:concl}

The multi-megabyte instruction footprints of contemporary server applications cause frequent BTB and L1-I misses, which have become major performance limiters. Because BTB capacity greatly affects front-end performance by dictating pipeline flush rate and the efficacy of fetch-directed instruction prefetching, commercial products allocate tens to hundreds of KBs of storage to BTBs. We observe that the single largest contributor to the BTB storage cost is the cost of storing branch target. We further observe that BTB storage cost can be drastically reduced by storing target offsets instead of full or even compressed targets. This is because targets of most branches lie relatively close to the branches themselves and our analysis shows that more than 99\% of offsets can be represented with at most half the bits required to store the full targets. Based on these observations, we propose a storage-effective BTB organization, called BTB-X, that stores target offsets in place of target address. Furthermore, BTB-X, an 8-way set associative BTB, uses differently sized ways with each storing offsets of a different length, thus accounting for the uneven distribution of offset lengths. Overall, BTB-X is capable of storing about 2.24x more branches than a conventional BTB and 1.3x more branches than a state-of-the-art BTB organization within the same storage budget.

%% file: 9_appendix.tex
\section{Artifact Appendix}
\label{sec:appendix}

\subsection{Abstract}
We implement BTB-X in Champsim simulator. Our artifacts provide the following: 1) BTB-X implementation in Champsim, 2) Link to workload traces, 3) Scripts for generating configuration files, launching simulations, and collecting results, and 4) Excel file for plotting the most important results. We identify three key results for artifact evaluation: a) Branch Target Offset distribution (Figure 4), b) BTB MPKI reduction (Figure 9), and c) Performance improvement (Figure 10).

\subsection{Meta-information}
\begin{itemize}
\item\runinsec{Compilation} Tested with GCC 8.5.0. It should also work with other recent GCC versions.

\item\runinsec{Code/Workloads} Download code/workloads from the provided link.

\item\runinsec{Experiments} Modify the provided scripts (as described below) to run simulations. 

\item\runinsec{Metrics} IPC, BTB MPKI, Branch Target Offset distribution.

\item\runinsec{Time needed to run experiments} Less than 30 minutes when running all traces in parallel.

\item\runinsec{Plotting graphs} Excel file, \emph{BTBX\_artifact\_results.xlsx}, is provided to plot graphs.
\end{itemize}

\subsection{Access to artifacts}

\begin{itemize}
\item\runinsec{Code} Download BTB-X implementation from~\cite{btbx-artifacts}. 

\item\runinsec{Workloads} The workloads can be downloaded from  

\url{https://drive.google.com/file/d/1qs8t8-YWc7lLoYbjbH\_d3lf1xdoYBznf/view?usp=sharing}

Place the workloads in \emph{\textless Path\_to\_code/dpc3\_traces/}.

\item\runinsec{Excel file} For plotting the graphs, download excel file, \emph{BTBX\_artifact\_results.xlsx}, from~\cite{btbx-artifacts}.

\end{itemize}

\subsection{System requirements}
Any hardware capable of running Champsim is sufficient. SLURM is recommended to run simulation on a cluster. The scripts are written in bash.

\subsection{Experiment workflow}

\begin{itemize}
\item\runinsec{Compilation} Champsim needs to be compiled with three BTB designs (convBTB, pdede, and BTBX) and two instruction prefetchers (no, fdip). Follow the instructions at~\cite{btbx-artifacts} to compile the code.

\runinsec{\textit{Important note on compilation}} IFETCH\_BUFFER needs to be 128 entries when compiling with “fdip” prefetcher and “FETCH\_WIDTH*2” entries when compiling with “no” prefetcher. This is because of how instruction fetch is implemented in baseline Champsim. IFETCH\_BUFFER size is defined in line 63 of \emph{\textless Path\_to\_code\textgreater/inc/ooo\_cpu.h}. 

\item\runinsec{Generating configuration files} 
Go to directory \emph{\textless Path\_to\_code\textgreater/launch/scripts/}. In the script file \emph{createConfig.sh}, point PATH\_TO\_CHAMPSIM to \textless Path\_to\_code\textgreater. Run this script (\emph{./createConfig.sh}) to generate config files needed by Champsim.

\item\runinsec{Running simulations}

\noindent \textit{Running all workloads:} Go to the directory \emph{\textless Path\_to\_code\textgreater/launch/}. In script file \emph{launch.sh}, replace the line \textless cluster\_launch\_command\_here\textgreater (line 64) with the command to run experiments on your cluster. A sample command is given that runs experiments on our cluster. Running this script (\emph{./launch.sh}) will run simulations, and the stats will be stored in directory \emph{\textless Path\_to\_code\textgreater/results\_50M/}.

\noindent \textit{Running a single workload:} An example command to run simulation for a single workload is provided at~\cite{btbx-artifacts}.

\end{itemize}

\subsection{Results}
\begin{itemize}
\item\runinsec{Collecting results}
Go to \emph{\textless Path\_to\_code\textgreater/collectStats/}. Run the script \emph{getResults.sh}, and it will collect results from all workloads and save them in a file \emph{all\_res}.

\item\runinsec{Plotting Results}
Download the \emph{all\_res} file. Open the provided excel file \emph{BTBX\_artifact\_results.xlsx}. Click on “Data” in MS-Excel top menu bar. Click on “Refresh All” in “Queries and Connections” ribbon, go to the folder where you stored \emph{all\_res} and double click on \emph{all\_res}. Now “Offset Distribution”, “MPKI”, and “Performance” sheets in the excel file should have plots for Figure 4, Figure 9, and Figure 10 respectively. 
\end{itemize}